\documentclass[fleqn,usenatbib]{mnras}

\usepackage{newtxtext,newtxmath}

\usepackage[T1]{fontenc}
\usepackage{ae,aecompl}

\usepackage{graphicx}	
\usepackage{amsmath}	
\usepackage{amssymb}	

\title[Barred galaxies in Auriga]{Chemodynamics of barred galaxies in cosmological simulations: On the Milky Way's quiescent merger history and in-situ bulge}

\author[Fragkoudi et al.]{F. Fragkoudi$^{1}$\thanks{E-mail:ffrag@mpa-garching.mpg.de}, 
R. J. J. Grand$^1$, R. Pakmor$^1$, G. Bl\'{a}zquez-Calero$^2$, I. Gargiulo$^{3,4}$,
\newauthor  F. Gomez$^{3,4}$, F. Marinacci$^5$, A. Monachesi$^{3,4}$, M. K. Ness$^6$, I. Perez$^{2,7}$, P. Tissera$^{8,9}$,
\newauthor S. D. M. White$^1$\\
\\
$^{1}$Max-Planck-Institut f\"{u}r Astrophysik, Karl-Schwarzschild-Str. 1, 85741 Garching, Germany \\
$^{2}$Departamento de F\'{i}sica Te\'{o}rica y del Cosmos, Universidad de Granada, Campus de Fuentenueva, E-18071 Granada, Spain \\
$^{3}$Instituto de Investigacion Multidisciplinar en Ciencia y Tecnolog\'{i}a, Universidad de La Serena, Raul Bitr\'{a}n 1305, La Serena, Chile\\
$^{4}$Departamento de Astronom\'ia, Universidad de La Serena, Av. Juan Cisternas 1200 Norte, La Serena, Chile\\
$^{5}$Department of Physics and Astronomy, University of Bologna, via Gobetti 93/2, I-40129 Bologna, Italy\\
$^{6}$Department of Astronomy, Columbia University, 550 West 120th Street New York, NY 10027, USA\\
$^{7}$Instituto Carlos I de F\'{i}sica Te\'{o}rica y Computacional, Universidad de Granada, E-18071 Granada, Spain\\
$^{8}$Departamento de Ciencias F\'{i}sicas, Universidad Andres Bello, Fernandez Concha 700, Santiago, Chile\\
$^{9}$Millennium Institute of Astrophysics, Fernandez Concha 700, Santiago, Chile\\
}

\date{Accepted XXX. Received YYY; in original form ZZZ}

\pubyear{2019}

\begin{document}

\defcitealias{Gargiuloetal2019}{G19}
\defcitealias{BlazquezCaleroetal2020}{B20}

\label{firstpage}
\pagerange{\pageref{firstpage}--\pageref{lastpage}}
\maketitle

\begin{abstract}
We explore the chemodynamical properties of a sample of barred galaxies in the Auriga magneto-hydrodynamical cosmological zoom-in simulations, which form boxy/peanut (b/p) bulges, and compare these to the Milky Way (MW).
We show that the Auriga galaxies which best reproduce the chemodynamical properties of stellar populations in the MW bulge have quiescent merger histories since redshift $z\sim3.5$: their last major merger occurs at $t_{\rm lookback}>12\,\rm Gyrs$, while subsequent mergers have a stellar mass ratio of $\leq$1:20, suggesting an upper limit of a few percent for the mass ratio of the recently proposed Gaia Sausage/Enceladus merger. These Auriga MW-analogues have a negligible fraction of ex-situ stars in the b/p region ($<1\%$), with flattened, thick disc-like metal-poor stellar populations. 
The average fraction of ex-situ stars in the central regions of all Auriga galaxies with b/p's is 3\% -- significantly lower than in those which do not host a b/p or a bar. While the central regions of these barred galaxies contain the oldest populations, they also have stars younger than 5\,Gyrs (>30\%) and exhibit X-shaped age and abundance distributions.
Examining the discs in our sample, we find that in some cases a star-forming ring forms around the bar, which alters the metallicity of the inner regions of the galaxy. Further out in the disc, bar-induced resonances lead to metal-rich ridges in the $V_{\phi}-r$ plane -- the longest of which is due to the Outer Lindblad Resonance. 
Our results suggest the Milky Way has an uncommonly quiet merger history, which leads to an essentially in-situ bulge, and highlight the significant effects the bar can have on the surrounding disc.
\end{abstract}




\section{Introduction}
Bars are common structures found in approximately two thirds of disc galaxies in the local Universe \citep{Eskridgeetal2000,Menendezetal2007,Aguerrietal2009,Gadotti2009,Mastersetal2011}, with this fraction decreasing towards higher redshifts, and reaching $\sim20\%$ at $z=1$ \citep{Shethetal2008,Melvinetal2014}, although a number of studies find evidence for the existence of bars at redshifts as high as $z\sim1.5-2$ (e.g. \citealt{Simmonsetal2014,Gadottietal2015}). 
They are known to affect their host galaxy in a variety of ways e.g. by pushing gas to the central regions, where it can form nuclear structures such as nuclear discs and rings (e.g. \citealt{Athanassoula1992b, Knapenetal2002, Comeronetal2010, Ellisonetal2011, Fragkoudietal2016, Sormanietal2018, deLorenzoCaceresetal2019,MendezAbreuetal2019,Leamanetal2019}; see reviews by \citealt{KormendyKennicutt2004} and \citealt{Athanassoula2013}).
Bars also re-shape the central regions of their host galaxy via the formation of a vertically extended bulge, often referred to as an X-shaped or boxy/peanut (b/p) bulge \citep{CombesSanders1981, Combesetal1990,Rahaetal1991,Patsisetal2002,Athanassoula2005,MartinezValpuestaetal2006,Quillenetal2014,Fragkoudietal2015}. 
In N-body simulations, b/p bulges form soon after the bar forms, either rapidly after a buckling instability (e.g. \citealt{Mihosetal1995,MartinezValpuestaetal2006}) or more slowly through resonant heating at the vertical inner Lindblad resonance of the bar \citep{PfennigerFriedli1991,Friedlietal1996,CeverinoKlypin2007,Quillenetal2014,Portailetal2015}. 

These and the aforementioned nuclear discs are sometimes collectively referred to as `pseudo-bulges', to differentiate them from dispersion-dominated, so-called `classical' bulges \citep{KormendyKennicutt2004}. To avoid confusion, we differentiate between b/p bulges -- which are formed by vertically extended orbits, and thus `puff-out' of the plane of the disc -- and nuclear discs or rings, which form out of gas pushed to the central regions by bars and which are flattened (disc-like) structures.
 Classical bulges are thought to form via violent processes such as dissipationless collapse, mergers or clump migration at high redshifts (e.g. \citealt{Eggenetal1962,Toomre1977, vanAlbada1982,Bournaudetal2007,NaabBurket2003,Hopkinsetal2009,Perezetal2013}; and the recent review by \citealt{BrooksChristensen2016}). The secular formation of b/p bulges and the violent formation mechanisms responsible for classical bulges leave different chemodynamical imprints, which can thus be used to decipher the formation history of their host galaxy.

The Milky Way (MW) is our closest barred galaxy, and therefore the bar's effects on its central regions and on its stellar disc can be explored in exceptional detail. 
There has been ample debate over the origin of the MW bulge -- whether a dispersion-dominated component formed from the dissipational collapse of gas or mergers, or a b/p bulge, formed via secular processes. Observations
in the near- and mid-infrared reveal that the MW bulge has a boxy or X-shape, pointing to its secular origin (e.g. \citealt{Dweketal1995,McWilliamandZoccali2010,Natafetal2010,Weggetal2013,NessLang2016}). However, a number of studies find the MW bulge to be an exclusively old population (e.g. \citealt{Zoccalietal2003,Clarksonetal2008,Valentietal2013}), with a negative radial metallicity gradient (e.g. \citealt{Zoccalietal2008}), which points to properties closer to those of a classical bulge. 
Further intensifying the debate, recent observational studies find that the MW bulge might not be exclusively old, with a significant fraction of stars younger than 8\,Gyrs \citep{Bensbyetal2013,Haywoodetal2016b,Bensbyetal2017}.
These seemingly contradictory properties have lent support to a hybrid scenario for the MW bulge, in which the metal-rich stellar populations are part of the b/p, formed from disc material, while the metal-poor populations constitute a separate dispersion-dominated, spheroidal, classical bulge component (e.g. see  \citealt{Babusiauxetal2010,RojasArriagadaetal2014,Barbuy2016, Barbuyetal2018} and references therein).  

On the other hand, our understanding of the disc of the Milky Way has also undergone a revolution of sorts, thanks to the recent second Gaia data release (Gaia DR2; \citealt{GaiaCollaboration2018}). Gaia DR2 has allowed for a detailed exploration of phase-space of the Milky Way's disc, revealing a number of previously unknown substructures, such as the Gaia snail or spiral \citep{Antojaetal2018}. Some of the most striking features the data have revealed are the prominent ridges in $V_{\phi}-r$ space \citep{Kawataetal2018,Antojaetal2018}, which have undulations in $V_r$ associated to them \citep{Fragkoudietal2019}. The bar has been proposed as a culprit for a number of these features including the observed ridges in the $V_{\phi}-r$ plane \citep{Fragkoudietal2019} and the Gaia spiral via the buckling instability (\citealt{Khoperskovetal2019}; but see \citealt{Laporteetal2019} for an alternative explanation). Furthermore, as shown recently by \citet{Khannaetal2019}, the ridges exhibit different abundance trends compared to phase space around them, which could perhaps give clues as to their origin. 

Additionally, recent  studies have probed the age and abundance structure of the inner disc of the Milky Way, finding seemingly contradictory results. On the one hand, \cite{Leungetal2019a,Leungetal2019b,Bovyetal2019} -- using APOGEE data in combination with machine learning techniques -- find that the bar of the Milky Way is metal-poor, while other studies such as \cite{Weggetal2019} -- using FLAMES \citep{Pasquinietal2000} spectra of red clump giant stars -- find that metal-rich stars in the inner disc tend to be on more elongated orbits, suggesting that the bar of the Milky Way is metal-rich. Furthermore, recent studies of local barred galaxies find that some bars tend to be more metal-rich than their surrounding disc, while others have similar or lower metallicities as compared to their surrounding disc population \citep{Neumannetal2020}. The aforementioned studies highlight the varied properties of barred galaxies, both for the Milky Way and external barred galaxies, as well as the tight interplay between the central regions of galaxies, the bar and the disc, all of which need to be explored in a unified framework within the global context of galaxy formation and evolution.

On the theoretical side, recent studies using tailored, isolated simulations of Milky Way-type galaxies, have shown that the metal-poor populations in the bulge of the Milky Way are in fact consistent with being composed of the thick disc seen at the Solar neighbourhood, with no need for an additional `classical' bulge component \citep{DiMatteo2016,Fragkoudietal2017b,Debattistaetal2017,Portailetal2017b,Haywoodetal2018,Fragkoudietal2018}. These models are able to explain the chemo-morphological and chemo-kinematic relations of stellar populations in the bulge \citep{Fragkoudietal2018,Gomezetal2018}, as well as its vertical and radial metallicity gradients \citep{Fragkoudietal2017c}.

While isolated simulations can be tailored to study specific galaxies in detail, such as the Milky Way, one would also like to be able to study the formation of bulges of MW-like galaxies in the full cosmological context.
Advances in resolution and physical fidelity (through sub-grid models) in recent cosmological zoom-in simulations have led to the formation of realistic disc galaxies, with smaller bulges (\citealt{Governatoetal2010,Bonolietal2016,BrooksChristensen2016}), which have thus started being used to study the properties of bars and b/p bulges in the context of the MW bulge (e.g. \citealt{Tisseraetal2018,Bucketal2018,Bucketal2019,Debattistaetal2019}). 
%
In general, however, these studies have explored single galaxies and therefore do not capture the diversity of formation histories that Milky Way mass galaxies can undergo. Also, while they reproduce a number of trends similar to the Milky Way bulge (such as e.g. morphology and global kinematic properties) they do not reproduce some of the key chemodynamical features of the Milky Way bar and bulge, such as the kinematical properties of the metal-poor (-1<[Fe/H]<-0.5) populations in the bulge (e.g. \citealt{Bucketal2019}), around which most of the debate about the origin of the MW bulge is centred.

We now have at our disposal for the first time a large sample of high resolution zoom-in cosmological simulations of Milky Way mass galaxies, the Auriga suite \citep{Grandetal2017,Grandetal2019}. These simulations develop realistic discs from diverse formation histories \citep{Gomezetal2017}, contain mostly bulges with low Sersic indices (\citealt{Gargiuloetal2019}, from now on G19) and develop bars and b/p bulges which at $z=0$ have structural properties in agreement with observations (\citealt{BlazquezCaleroetal2020}, from now B20). We can therefore now study the formation of bars and b/p's in the full cosmological context, exploring the chemodynamical imprints left by their formation history. This allows us to constrain the merger history of the Milky Way (see also \citealt{Monachesietal2019}), and to explore consistency with the recently proposed Gaia Sausage/Enceladus merger \citep{Belokurovetal2018,Haywoodetal2018b,Helmietal2018}. As we will show, these models are able to reproduce a number of chemodynamical properties of the MW bulge, thus shedding light on its formation history, while also allowing us to explore the effects of the bar on the disc, not only in terms of kinematics but also by taking into account the chemical enrichment and ages of stellar populations in the disc. 

This paper is the first in a series exploring the properties of bars in the Auriga cosmological simulations. Here we explore the chemodynamical properties of Auriga galaxies with prominent b/p bulges, comparing them to the MW bulge and connecting them to their assembly history. We also explore the effects that bars have on the discs of MW-type galaxies. 
The paper is structured as follows: in Section \ref{sec:auriga} we describe the Auriga simulations, focusing on the sample studied here, and show some statistical properties of barred galaxies in Auriga. In Section \ref{sec:ageabund} we describe the age and abundance distributions in our sample, focusing on the bar-b/p region. In Sections \ref{sec:chemorph} and \ref{sec:chemokinem} we describe the chemo-morphological and chemo-kinematic relations of stellar populations in the central regions and then compare them to the Milky Way bulge, while in Section \ref{sec:SFH} we relate these to the galaxies' merger history and fraction of ex-situ stars. In Section \ref{sec:bareffectdisc} we explore the effects of the bar on the inner and outer disc. In Section \ref{sec:discussion} we discuss some of the implications of our findings in terms of the inner disc of the Milky Way, and in Section \ref{sec:summary} we conclude and summarise our results.

\section{The Auriga Simulations}
\label{sec:auriga}
\begin{figure*}
\centering
\includegraphics[width=0.95\textwidth]{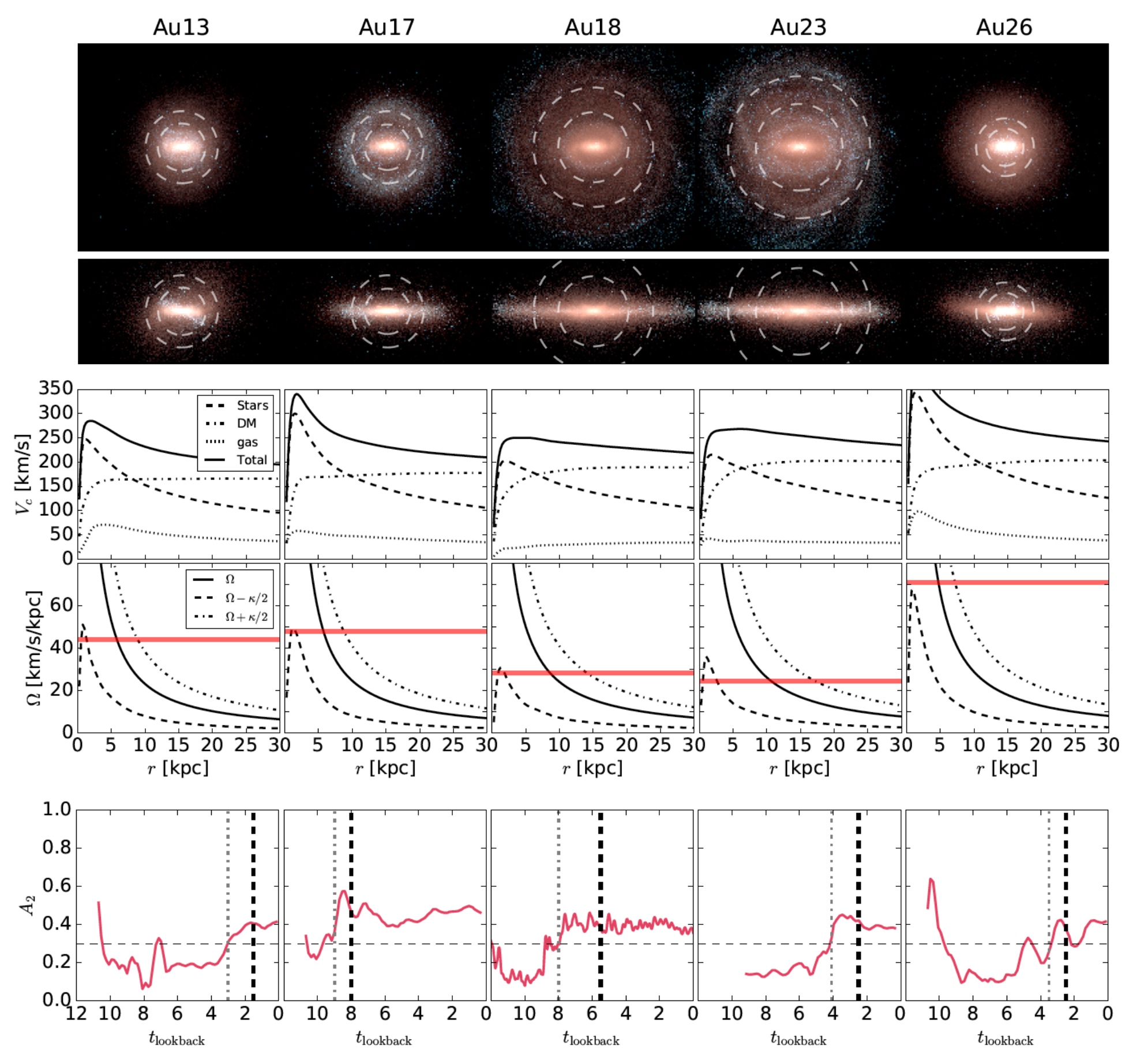}
\caption{Properties of the five Auriga galaxies explored in this study. \emph{Top row:} RGB images -- synthesized from a projection of the K-, B- and U-band luminosity of stars -- of the subsample of Auriga galaxies which we explore in this study. The size of the face-on panels is $50\times50\,\rm kpc$ and of the edge-on panels $50\times 25 \, \rm kpc$ (for edge-on projections the line-of-sight is along the bar minor axis). The halo number is denoted at the top of each plot and the inner and outer dashed circles show the corotation and Outer Lindblad Resonance radius respectively. \emph{Second row:} Circular velocity profiles for the sample of galaxies: total (solid lines), stellar component (dashed lines), dark matter component (dot-dashed lines) and gaseous component (dashed lines). \emph{Third row:} Angular frequency plots showing $\Omega$ the angular frequency (solid) and $\Omega - \kappa/2$ and $\Omega + \kappa/2$ in dashed and dot-dashed respectively, where $\kappa$ indicates the radial frequency of stars. The horizontal red line denotes the pattern speed of the bar. \emph{Fourth row:} Bar strength $A_2$ as a function of lookback time. The vertical dot-dashed line marks the formation of a strong bar and the thick dashed line the formation of the b/p.} 
\label{fig:rgball}
\end{figure*}

\begin{figure}
\centering
\includegraphics[width=0.47\textwidth]{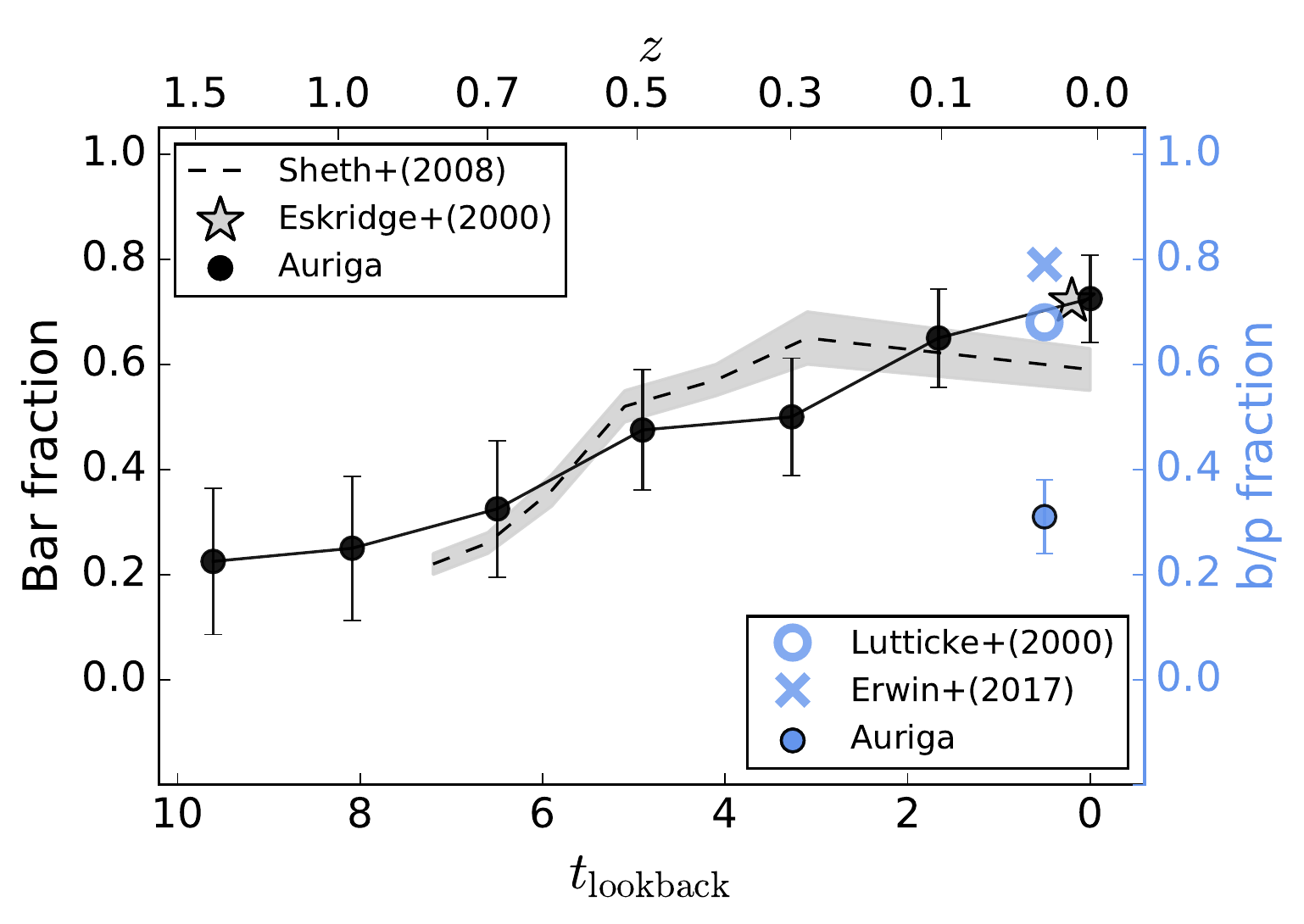}
\includegraphics[width=0.42\textwidth]{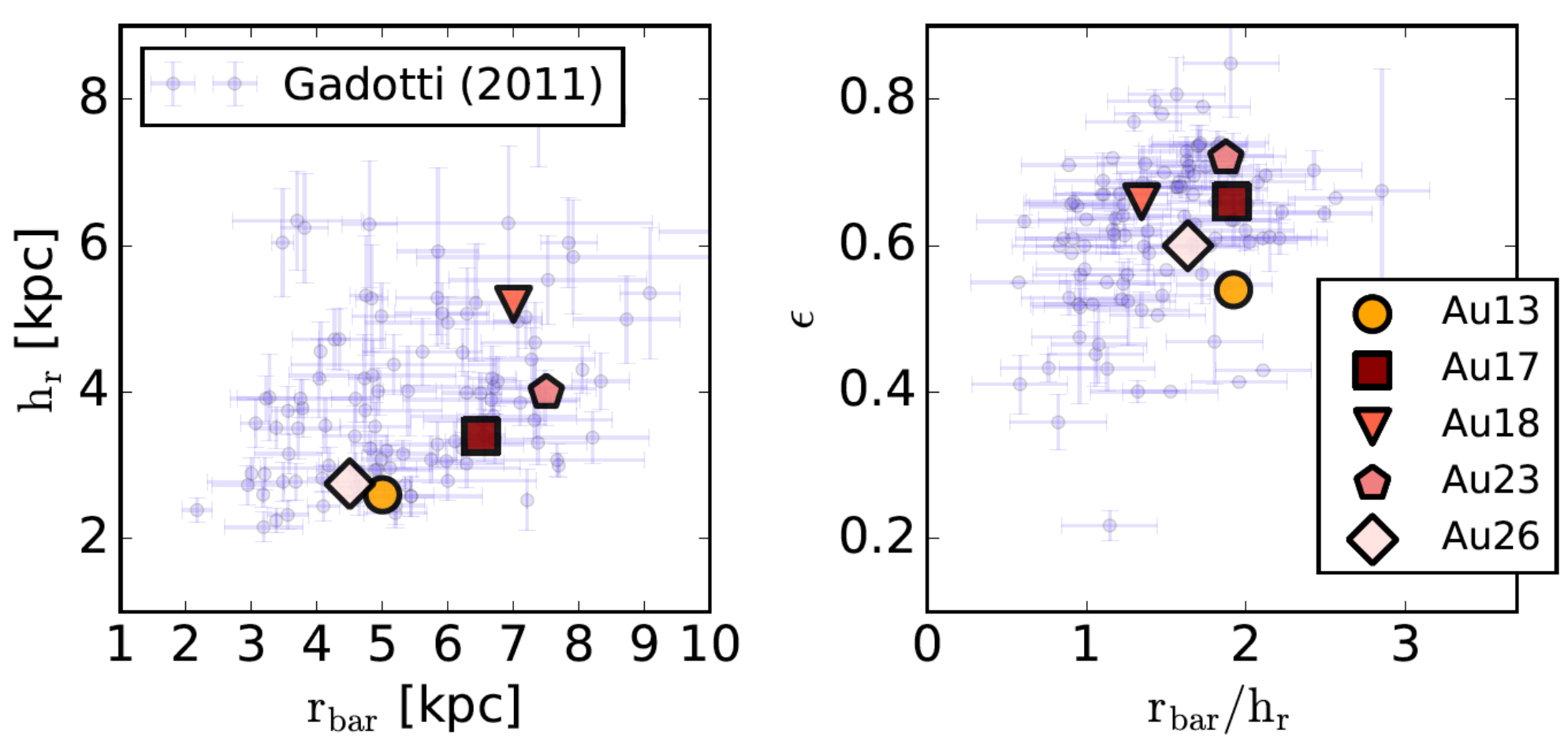}
\caption{Statistical properties of the galaxies in the entire Auriga sample and in the subsample examined here. \emph{Top:} Bar fraction as a function of redshift for the entire suite of Auriga simulations, compared to observations \citep{Shethetal2008,Eskridgeetal2000}. The blue points indicate the fraction of b/p bulges in barred galaxies in Auriga, comparing to observations from \citet{Luttickeetal2000} and \citet{ErwinDebattista2017} (right $y$-axis). \emph{Bottom Left:} Disc scale-length vs bar length for SDSS galaxies from Gadotti (2011) (blue circles) and for the five Auriga galaxies explored in this study (symbols). \emph{Bottom Right:} Corrected bar ellipticity (see text) vs bar length over disc scale-length.} 
\label{fig:barfrac}
\end{figure}

The Auriga simulations  \citep{Grandetal2017,Grandetal2019} are a suite of cosmological magneto-hydrodynamical zoom simulations of haloes with masses in the range of $0.5 \times 10^{12}-2 \times 10^{12}\rm M_{\odot}$ which run from redshift $z=127$ to $z=0$ with cosmological parameters: $\Omega_{\rm m}=0.307$, $\Omega_{\rm b}=0.048$ and $\Omega_{\rm \Lambda}=0.693$, and a Hubble constant of $H_{\rm 0} =100h\, \rm km\,s^{-1}\,\rm Mpc^{-1}$,where $h=0.6777$ \citep{Planck2014XVI}. 
The simulations are performed with the magnetohydrodynamic code {\sc AREPO} \citep{Springel2010,Pakmoretal2016}, with a comprehensive galaxy formation model (see \citealt{Vogelsbergeretal2013, Marinaccietal2014a,Grandetal2017}, for more details) which includes primordial and metal line cooling, a prescription for a uniform background ultraviolet field for reionization (completed at $z=6$), a subgrid model for star formation, stellar evolution and feedback, magnetic fields, and black hole seeding, accretion and feedback. 
The dark matter particles have  a mass of $\sim 4\times10^5 \rm M_{\odot}$ and the stars and gas have a mass resolution $\sim 5\times10^4 \rm M_{\odot}$. The physical softening of collisionless particles grows with time and corresponds to a fixed comoving softening length of 500\,$h^{-1}$pc, while the maximum physical softening allowed is 369\,pc (see \citealt{Poweretal2003} for reasonable softening parameters). The physical softening for the gas cells is scaled by the gas cell radius with a minimum limit of the softening equal to that of the collisionless particles.

Star formation and stellar feedback is modelled as follows: if a given gas cell is eligible for star formation, it is converted (according to the \citealt{Chabrier03} initial mass function) either into a star particle -- in which case it represents a single stellar population of a given mass, age and metallicity -- or into a site for SNII feedback. 
In the latter case, this particle is launched in a random direction as a  wind particle with a velocity that scales with the 1-D local dark matter velocity dispersion (see \citealt{Grandetal2017} for more details). Its metal content is determined by the initial metallicity of the gas cell from which the wind particle originated, i.e. it is loaded with $\eta=0.6$ of the total metals of the parent gas cell. For the stellar particles, we model the mass loss and metal enrichment from SNIa and AGB stars by calculating the mass moving off the main sequence for each star particle at each timestep. The mass and metals are then distributed among nearby gas cells with a top-hat kernel.
We track a total of nine elements: H, He, C, O, N, Ne, Mg, Si, and Fe and in what follows we use (Mg+Si+O)/3 to study the $\alpha$-abundances.

The simulations form disc-dominated star-forming galaxies with flat rotation curves that reproduce a range of observed scaling relations such as the Tully-Fisher relation \citep{Grandetal2017} and the size-mass relation of HI gas discs \citep{Marinaccietal2017}. They also form instabilities in the discs such as bars and boxy/peanuts which have structural properties similar to those of observed bars \citepalias{BlazquezCaleroetal2020} and mainly consist of so-called pseudo-bulges \citepalias{Gargiuloetal2019}, reproducing what is found for disc galaxies in the local Universe \citep{KormendyKennicutt2004,Gadotti2009}. 
Four of the haloes used in this study (Au13, Au17, Au23 and Au26) are the original haloes presented in \cite{Grandetal2017}, while Au18 is a re-run of the original halo 18 from \citet{Grandetal2017}, for which we have high cadence snapshot outputs, saved every 5\,Myr\footnote{While the initial conditions of the halo are the same as those of the original halo in \cite{Grandetal2017}, the final galaxy is not identical due to differences in the integration time-step. However the overall properties of the galaxy and its bar are broadly similar as a function of redshift, which gives confidence that the properties of strongly barred galaxies are to some extent robust.}.   

\subsection{Analysis}
\label{sec:analysis}
To obtain the bar strength in our sample of simulated galaxies we select the stellar particles in the disc and calculate the Fourier modes of the surface density as,
\begin{equation}
a_m(R) = \sum^{N}_{i=0}  {\rm m_i} \cos(m\theta_i), \,\,\,\,\, \,\,\,\,\, m=0,1,2,...,
\end{equation}
\begin{equation}
b_m(R) = \sum^{N}_{i=0} {\rm m_i} \sin(m\theta_i),  \,\,\,\,\,\,\,\,\,\, m=0,1,2,...
\end{equation}

\noindent where $\rm m_i$ is the mass of particle $i$, $R$ is the cylindrical radius, $N$ is the total number of particles in that radius and $\theta$ is the azimuthal angle.
To obtain a single value for the bar strength we take the maximum of the relative $m=2$ component within the inner 10\,kpc as,
\begin{equation}
A_2 = \rm max \frac{\sqrt{ \left( a^2_2 + b^2_2 \right)} }{a_0}.
\end{equation} 
Depending on the analysis, this can be calculated for all stars in the disc, or for each mono-age or mono-abundance population separately. In what follows we define a (strong) bar as having formed when $A_2>0.3$. We always also visually inspect the bars to be sure that large values of $A_2$ are not due to transient effects, such as an off-centering due to a merger etc.

The bar pattern speed in our fiducial model, Au18, is obtained by calculating the $m=2$ phase in each snapshot and then calculating the bar pattern speed $\Omega_p$, 
\begin{equation}
\Omega_p = \frac{\Delta \theta}{\Delta t}.
\end{equation}
For the other four haloes investigated in this study, for which we do not have high enough cadence outputs in order to calculate the bar pattern speed directly from the temporal evolution of the simulations, we calculate $\Omega_{\rm p}$ using the Tremaine-Weinberg method (TW; \citealt{TremaineWeinberg1984}). The method relies on the continuity equation and on the disc having a well defined pattern speed, such as the bar pattern speed $\Omega_{\rm p}$, and can be easily used to calculate the pattern speed using slits placed perpendicular to the line of nodes of the galaxy (see \citealt{TremaineWeinberg1984} for more details). We first tested the TW method on Au18, the fiducial model, and found the parameters such as bar orientation, disc inclination and number of slits and their extent along the bar major axis etc., that give the most accurate results (see \citealt{Debattista2003,GarmaOehmichenetal2019} and references therein for tests of the TW method). To calculate the pattern speeds in the four galaxies in our sample we employ an inclination angle of the disc of $i=45$\,degrees and rotate the bar such that it has an angle of 60\,deg with respect to the line of nodes. We tested our implementation of the TW method on other reruns of the Auriga sample for which we have high cadence outputs (and which will be presented in future work; Fragkoudi et al. in prep.) and found that we can recover the true pattern speed with an accuracy of 5\%.


\subsection{Barred-boxy/peanut sample}
\label{sec:bpsample}
In this study, we focus on five halos from the Auriga suite, shown in Figure \ref{fig:rgball}, which have bars and prominent b/p bulges which are readily identified in their edge-on\footnote{In what follows, unless explicitly stated, when referring to edge-on projections we project the galaxy along the $y$-axis, with the bar's semi-major axis aligned with the $x$-axis.} surface density projections. We consider a galaxy to have a b/p bulge when the X-shape of the bar is visible in the edge-on projection along with a `bump' in the mean height of stars as a function of radius (see Figure \ref{fig:peanut_strength}). 
Four other Auriga barred galaxies show hints of a b/p bulge in the process of forming, which we term `weak b/p's' (these b/p's are identified because there is a `bump' in the mean height of \emph{young} stars). However, as these are too weak to be seen in the edge-on projection of \emph{all} stars they would likely not be identified as peanut galaxies observationally, therefore we do not include them in this study\footnote{We note that there is no strict definition of how to classify a peanut (see also \citealt{CiamburGraham2016}). In \citetalias{BlazquezCaleroetal2020}, we use un-sharp masking of edge-on projections to identify b/p's with which we identify 6 b/p's, five of which are the prominent ones presented in this study; the sixth one has a very weak b/p which we include in our sample of `weak b/p's'.}.  

We note that the  fraction of barred galaxies in the entire Auriga sample -- 40 haloes presented in \citet{Grandetal2017,Grandetal2019} -- as a function of redshift is consistent with observations (e.g. \citealt{Shethetal2008}): i.e. we find that $\sim70\%$ of the Auriga galaxies have bars at $z=0$ with the fraction steadily decreasing towards higher redshifts, and reaching a plateau of $\sim20\%$ at $z=1.5$ -- see Figure \ref{fig:barfrac}. We also compare the fraction of b/p's in Auriga (including weak b/p's) to observed fractions of b/p's in the local Universe \citep{Luttickeetal2000,ErwinDebattista2017}; the fraction of b/p's in Auriga is low (30\%) -- and is even lower if we exclude the weak b/p's which would be hard to detect observationally  -- compared to the observed fraction of $\sim70\%$ (and see also \citetalias{BlazquezCaleroetal2020}). This could be due to the slightly too hot discs in our simulations (e.g. see \citealt{Grandetal2016}); we will discuss this and the formation of barred/peanut galaxies in the cosmological context in more detail in upcoming work (Fragkoudi et al. in prep.).

In the false-colour face-on and edge-on RGB images of our sample of galaxies in Figure \ref{fig:rgball}, we see the overall morphology of the Auriga galaxies in our sample, with the corotation radius (CR) and Outer Lindblad Resonance (OLR) of the bar marked with the inner and outer and dashed lines respectively. The galaxies have interesting morphological features, similar to many barred galaxies in the local Universe; for example haloes Au17, Au18 and Au23 have a red and quenched region inside the bar radius (often referred to as the `star formation desert', e.g. \citealt{Jamesetal2009}), and a blue star forming disc. On the other hand, haloes Au13 and Au26 have ongoing star formation in the central regions. We also note the presence of a star forming inner ring inside the CR, which surrounds the bar, in haloes Au18 and Au23. These and other ring like structures are a common feature of barred galaxies, and have been typically thought to form due to gas piling up at bar-induced resonances (\citealt{ButaCombes1996}; and see Section \ref{sec:effectring} for more discussion on these rings).

In the third row of Figure \ref{fig:rgball} we plot the rotation curves for these galaxies\footnote{Obtained by approximating the mass distribution as spherical and using $V_c(r) = \sqrt{GM(<r)/r}$}. Haloes Au18 and Au23 exhibit almost flat outer profiles, while Au13, Au17 and Au26 have rather peaked profiles in the central regions due to a more concentrated stellar distribution (we note that all the galaxies in our sample have slightly declining rotation curves, as expected for massive spiral galaxies -- e.g. \citealt{SofueRubin2001}). In the third row of Figure \ref{fig:rgball} we show the angular frequency curves, as well as the $\Omega \pm \kappa/2$ curves, where $\kappa$ is the radial frequency of stars on near circular orbits\footnote{$\kappa(R_{\rm g}) = \sqrt{ (R\frac{\rm d\Omega^2}{\rm d R} + 4\Omega^2)}$ in the epicyclic approximation; see \citet{BT2008}}. The bar pattern speed $\Omega_p$ is indicated with the horizontal red line, while its intersection with the $\Omega$ and $\Omega \pm \kappa/2$ curves gives the approximate locations of the CR and the Outer and Inner Lindblad Resonances respectively\footnote{These are the locations of the resonances strictly only for mildly non-axisymmetric systems}. 
In the fourth row of Figure \ref{fig:rgball} we show the evolution of bar strength as a function of time; we mark the bar formation time (i.e. for $A_2>0.3$) with a thin dot-dashed vertical line and the formation of the b/p with a thick dashed line. 

The properties of the bars in the Auriga simulations at $z=0$ are in general in good agreement with those of observed galaxies, as can be seen in the bottom panels of Figure \ref{fig:barfrac}, where we show the relation between disc scale-length, $\rm h_r$, and bar length, $\rm r_{bar}$, as well as bar ellipticity, $\epsilon$, vs $\rm r_{bar}/\rm h_r$ (see also \citetalias{BlazquezCaleroetal2020} who carry out a detailed comparison of structural properties of bars and b/p's at $z=0$ in Auriga with observations). Here we derive the disc scale-lengths by fitting the 1D surface density with a disc and bulge component. The bar lengths are obtained from ellipse fitting the surface density images where the bar length is derived as the minimum between the first minimum of ellipticity profile or when the angle of the ellipses changes by more than 5 degrees (see \citealt{Erwin2005}). The bar ellipticity is obtained as the maximum ellipticity of our fits in the bar region\footnote{\cite{Gadotti2008} showed that the ellipticity obtained using ellipse fits is 20\% lower than that obtained using 2D image decompositions (see their Section 3.4); we therefore correct our ellipticities accordingly in order to be able to compare with the observed sample.}. We see that the bars in our simulations match well the properties of observed barred galaxies in \cite{Gadotti2011} in terms of bar length, ellipticity and disc scalelength. 

In what follows we refer to Au18 as our fiducial model; this run has high cadence outputs and is therefore used for tests in much of the analysis that follows, while furthermore, as we will show in the next Sections, the model has similar chemodynamical properties to the Milky Way bulge.

\section{Ages \& abundances in bars and b/p bulges}
\label{sec:ageabund}
\begin{figure*}
\centering
\includegraphics[width=0.95\textwidth]{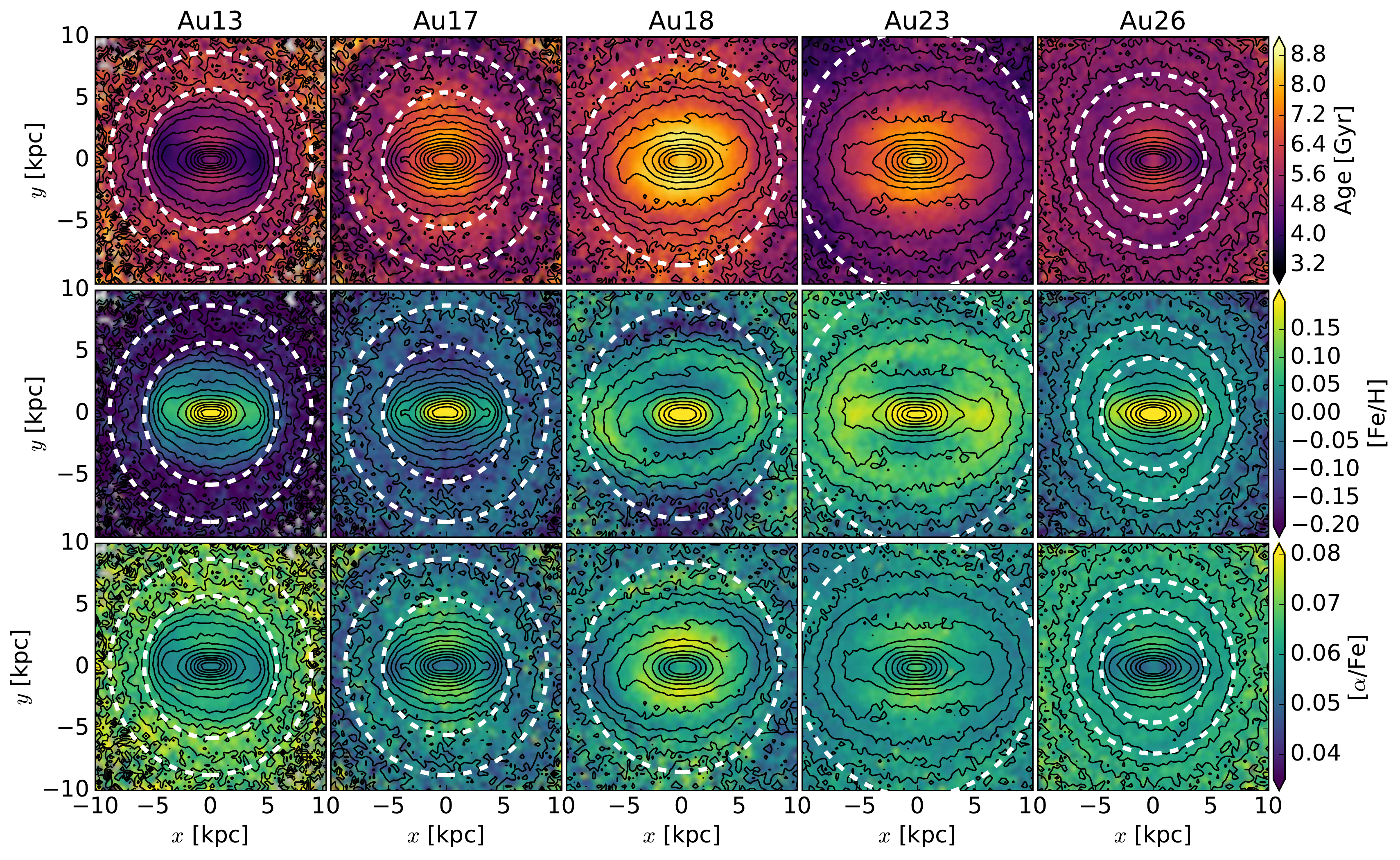}
\caption{Face-on projection of the ages and abundances of stars within $|z|<0.5\,\rm kpc$ in our sample of galaxies. \emph{Top row:} Mass-weighted mean age. \emph{Second row:} Mass-weighted mean [Fe/H]. \emph{Third row:} Mass-weighted mean [$\alpha$/Fe] distribution. In all panels the inner and outer dashed circles denote the CR and OLR radius. The black curves indicate iso-density contours.} 
\label{fig:xyall_ages}
\end{figure*}


\begin{figure*}
\centering
\includegraphics[width=0.95\textwidth]{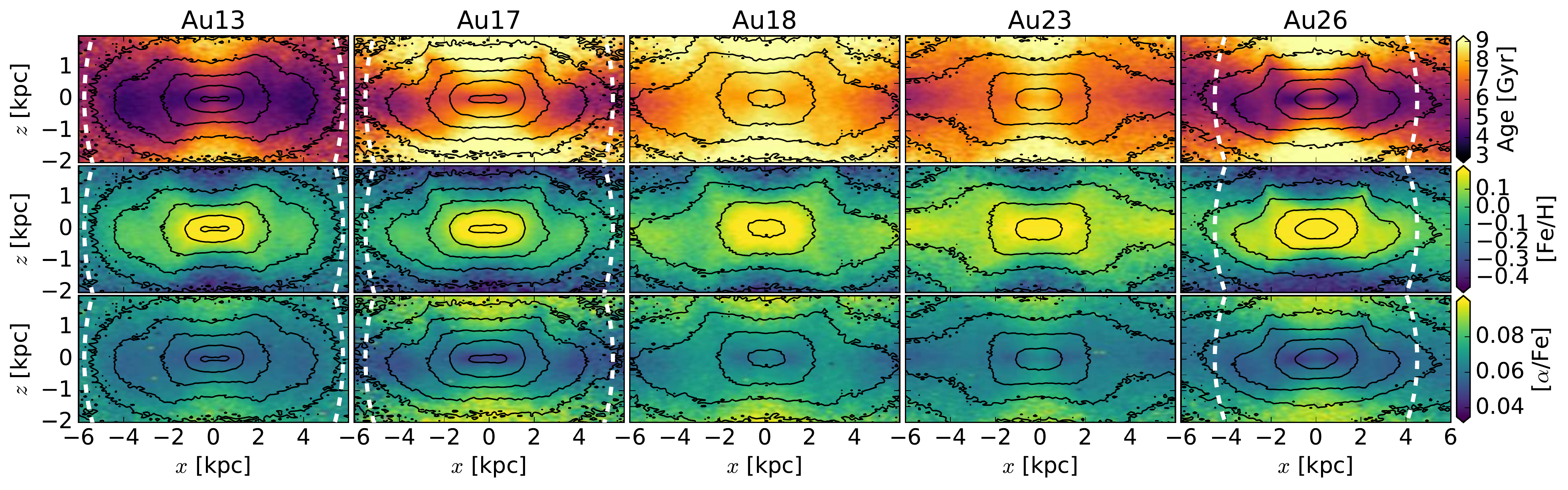}
\caption{Ages and abundances of the boxy/peanut bulges in our sample: \emph{First row:} Mass-weighted mean age in the boxy/peanut. \emph{Second row:} Mass-weighted mean [Fe/H]. \emph{Third row:} Mass-weighted mean [$\alpha$/Fe]. The vertical dashed lines indicate the corotation radius (for the cases where it falls inside the plotted region). We see that the edges of the b/p bulges are traced by younger, more metal-rich and more $\alpha$-poor populations. } 
\label{fig:xy_edgeon_feh_all}
\end{figure*}

\begin{figure}
\centering
\includegraphics[width=0.23\textwidth]{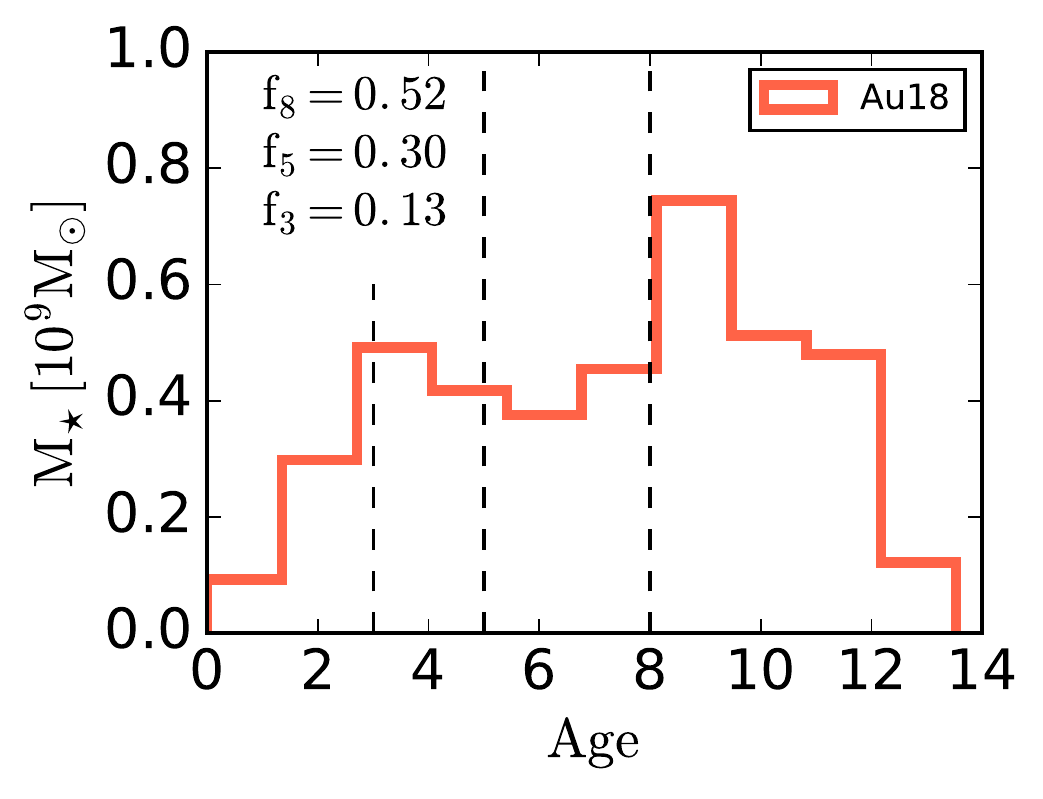}
\includegraphics[width=0.23\textwidth]{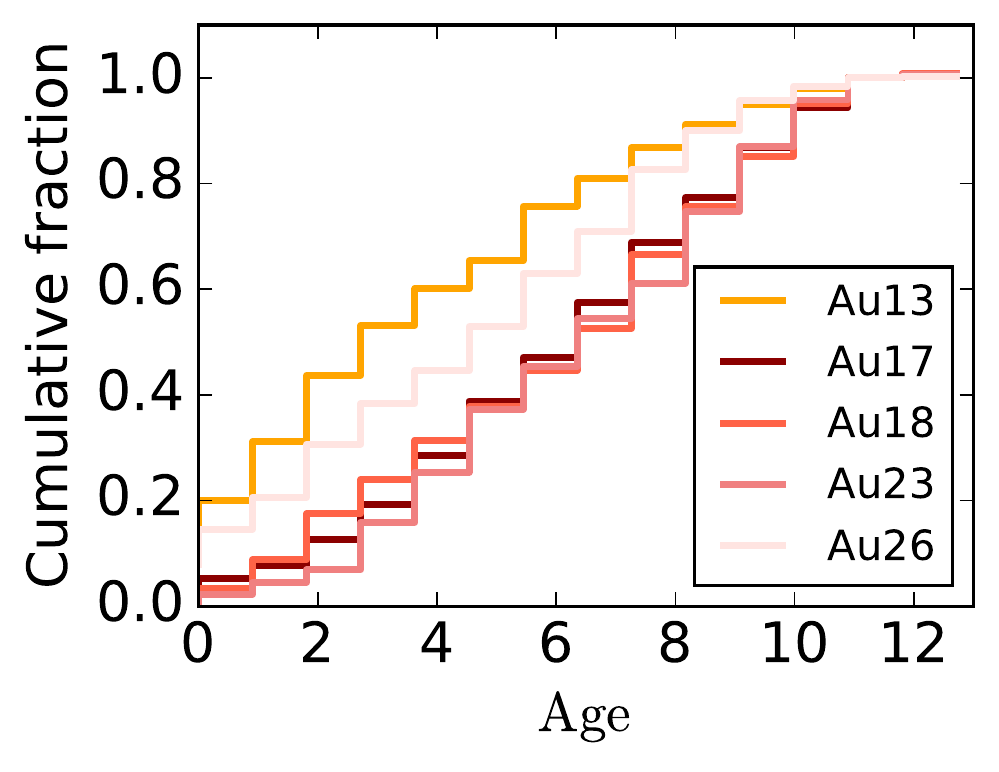}
\caption{Age distributions in the inner regions of the Auriga galaxies explored in this study. \emph{Left panel:} Distribution of ages inside the boxy/peanut bulge of halo Au18. In the top left corner we denote the fraction of stars younger than a certain age (<8, 5, 3\,Gyrs). \emph{Right panel:} Cumulative ages inside the boxy/peanut bulges of all the haloes in our sample.} 
\label{fig:cumul}
\end{figure}

\subsection{Face-on projection}
In Figure \ref{fig:xyall_ages} we show face-on maps of mass-weighted mean age, metallicity and $\alpha$-abundances for all galaxies in our sample, selecting stars within $|z|<0.5\,\rm kpc$ from the plane of the galaxy. We see that there is a large variety in the mean ages and abundances of the bars and discs in the sample. In haloes Au17, Au18 and Au23 the bar region is overall old (mean age $>8\,\rm Gyrs$), while haloes Au13 and Au26 have more recent episodes of star formation (as we will see below in Section \ref{sec:SFH}) and therefore have younger ages ($\sim4\,\rm Gyr$) in the bar region. 

We see that in all cases the ends of the bar tend to have younger ages than stars found perpendicular to the bar. If we were to trace the mean age of stars in cuts perpendicular to the bar, we would therefore find a decreasing age gradient towards the center of the bar, with younger ages clustering along the bar semi-major axis (and see also \citealt{Wozniak2007}). This is likely a consequence of the kinematic differentiation of stars in the bar (which we will discuss in the next Section in more detail) where younger populations in the bar have more elongated shapes than older populations which are rounder (see \citealt{Debattistaetal2017,Fragkoudietal2017b}). The variation in morphology for populations of different ages naturally leads to such an age gradient perpendicular to the bar. This behaviour has been recently observed in local barred galaxies using spectroscopic data from the MUSE-TIMER survey -- see \cite{Neumannetal2020} for more details.
\vspace{1cm}

In the second row of Figure \ref{fig:xyall_ages} we show the metallicity distribution in our sample. We see that in all haloes, the bar is more metal-rich than the surrounding disc, with the exception of haloes Au18 and Au23 where a prominent inner ring is formed, which is star-forming and metal-rich (see Section \ref{sec:effectring} for a more detailed discussion on these inner rings). In these two cases (Au18 and Au23) only the inner 1.5\,kpc of the galaxy is more metal-rich than the inner ring, while along the bar the metallicity is lower than that of the inner ring. The metal-rich inner-most regions of the bars in our sample, i.e. inside $\sim$1.5\,kpc, perhaps indicate the formation of nuclear discs in the Auriga galaxies. However these would be larger than those found in observed galaxies (here of the order of 1-2\,kpc while nuclear discs tend to have sizes of the order of a few hundred parsec, e.g. \citealt{Comeronetal2010}) which could be due to resolution issues in the central-most kiloparsec or possibly due to the AGN feedback implementation; this will be the subject of future investigations. We also see that, especially in the region of the bar, there are clear azimuthal variations in the metallicity maps, with metallicity gradients being flatter along the bar than perpendicular to it.

In the third row of Figure \ref{fig:xyall_ages} we show the $\alpha$-abundances in the discs of our haloes. We see that the inner regions are on average more $\alpha$-enhanced, with a similar gradient as for the age, i.e. $\alpha$-poor stars are concentrated along the bar major axis. We also see that for haloes Au18 and Au23, the aforementioned possible metal-rich nuclear discs, correspond to regions of low $\alpha$-enhancement, as is expected for these types of inner structures which form through secular processes.

\subsection{Edge-on projection}

The MW bulge has long been thought to be exclusively old, as found by studies using colour-magnitude diagrams in fields towards the bulge (e.g. \citealt{Zoccalietal2003,Clarksonetal2008,Valentietal2013}). On the other hand, recent studies such as those of \cite{Bensbyetal2013,Bensbyetal2017}, which derive the ages of microlensed stars in the bulge, find that there is in fact a wide distribution of ages in the bulge, with up to 50\% of metal-rich ([Fe/H]>0) stars younger than 8\,Gyrs. Furthermore, \cite{Haywoodetal2016} recently re-analysed the CMD which was used in \cite{Clarksonetal2008} and found that when allowing for an evolving Age-[Fe/H] relation for stars in the bulge, the bulge CMD is better fit by isochrones with a spread of ages. They furthermore found that all stars with [Fe/H]>0 can be younger than 8\,Gyrs (which would make up $\sim50\%$ of all the stars in the bulge). 

In what follows we analyse the age distributions for the b/p's in our sample of simulated galaxies.
In Figure \ref{fig:xy_edgeon_feh_all} we show edge-on maps of mean ages, metallicities and abundances in our sample (to remove contamination from disc stars we exclude stars outside galactocentric radius $R=6\,\rm kpc$). We see that, as in the face-on distribution of ages in the galaxies in Figure \ref{fig:xyall_ages}, haloes Au17, Au18 and Au23 have on average older b/p bulges, while Au13 and Au26 have younger b/p's, since they have recent ongoing star formation in the central regions. In all haloes we see an X-shaped distribution of ages, with the younger populations dominating the X-shape of the peanut -- i.e. the relative fraction of young to old stars will depend on which region of the bulge is explored. 
In the second and third rows of the figure we show the edge-on metallicity and $\alpha$-abundance distributions. We see that all the b/p bulges demonstrate a pinched X-shape metallicity and $\alpha$-abundance distributions (see also e.g. \citealt{Gonzalezetal2017,Debattistaetal2017,Fragkoudietal2018}). 

We examine the age distribution of stars in the entire b/p bulge region (here we restrict this cut to $R<4\, \rm kpc$ and $|z|<2\, \rm kpc$ to take only stars within the central-most regions) of our fiducial model, Au18, in the left panel of Figure \ref{fig:cumul}. We see that there is a significant fraction of young stars in the boxy/peanut bulge region, with 52\% of stars younger than 8\,Gyr, 30\% of stars with Ages<$5\,$Gyrs and 13\% of stars with Ages$< 3\,$Gyrs. In the right panel of Figure \ref{fig:cumul} we show the cumulative fraction of ages in all five haloes in our sample. We see that all haloes show significant fractions of young stars inside the boxy/peanut bulge, with stars younger than 5\,Gyrs ranging between 25-60\% depending on the halo and its star formation history\footnote{We note that, as we discuss in later sections, Au18 has tentatively a similar star formation history as the MW}. It is worth noting that for none of our b/p bulges are the ages exclusively old (i.e. all older than 10\,Gyr). 
Our models therefore suggest that there is a spread of ages in b/p's that are formed in the full cosmological setting, with a non-negligible fraction of young stars in the central regions of Milky Way-mass galaxies.

\section{Morphological properties of stellar populations in bars and b/p bulges}
\label{sec:chemorph}

\begin{figure*}
\centering
\includegraphics[width=0.8\textwidth]{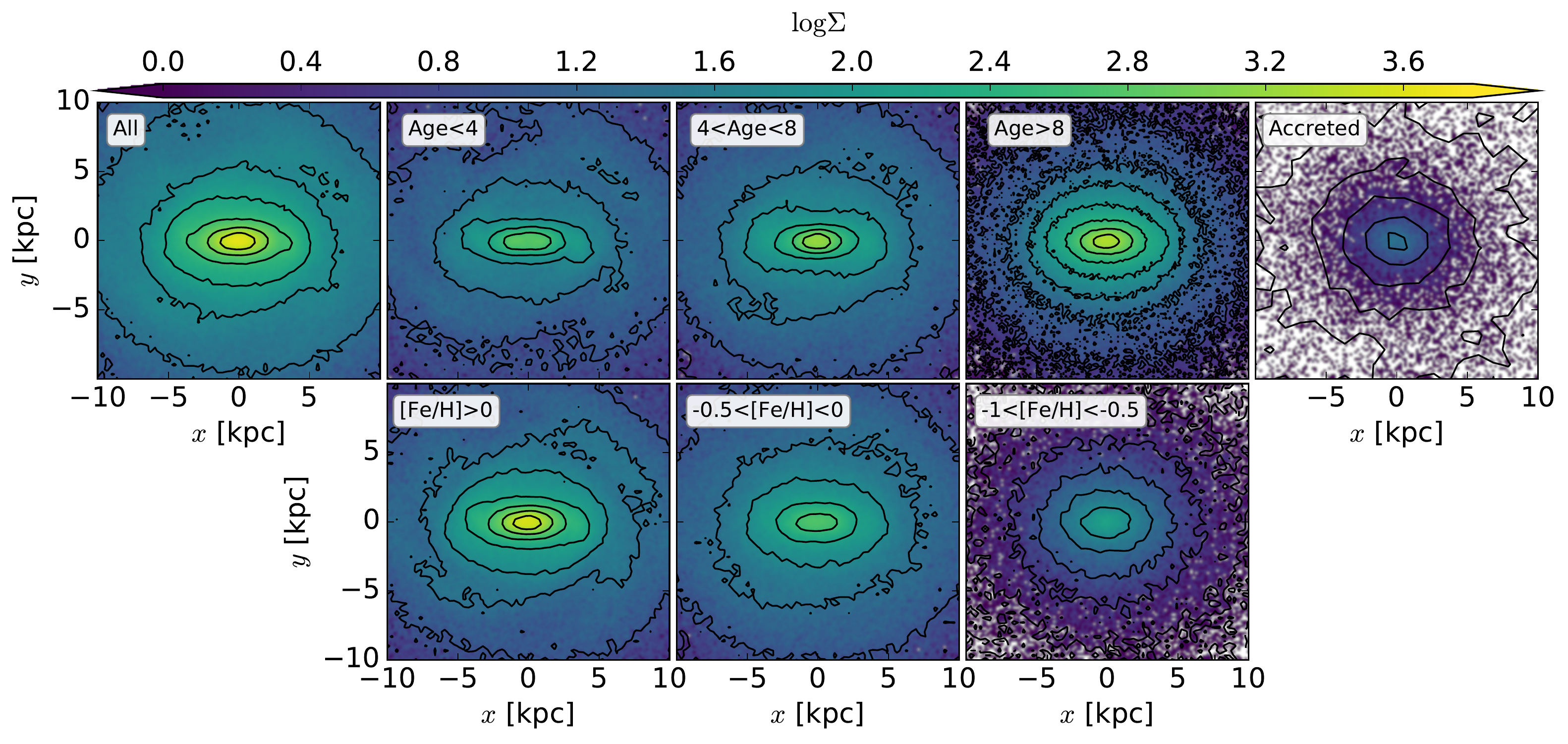}
\caption{Chemo- and chrono-morphological properties of Au18: \emph{Top:} Face-on surface density projection of Au18 for all stars (left) and for stars with different ages (subsequent columns; age is denoted in the upper left corner of each panel) as well as accreted stars (right panel). \emph{Bottom:} Face-on surface density projection for stars with different metallicities, as denoted in the top left corner of each panel.}
\label{fig:xy_faceon_edgeon}
\end{figure*}

\begin{figure*}
\centering
\includegraphics[width=0.8\textwidth]{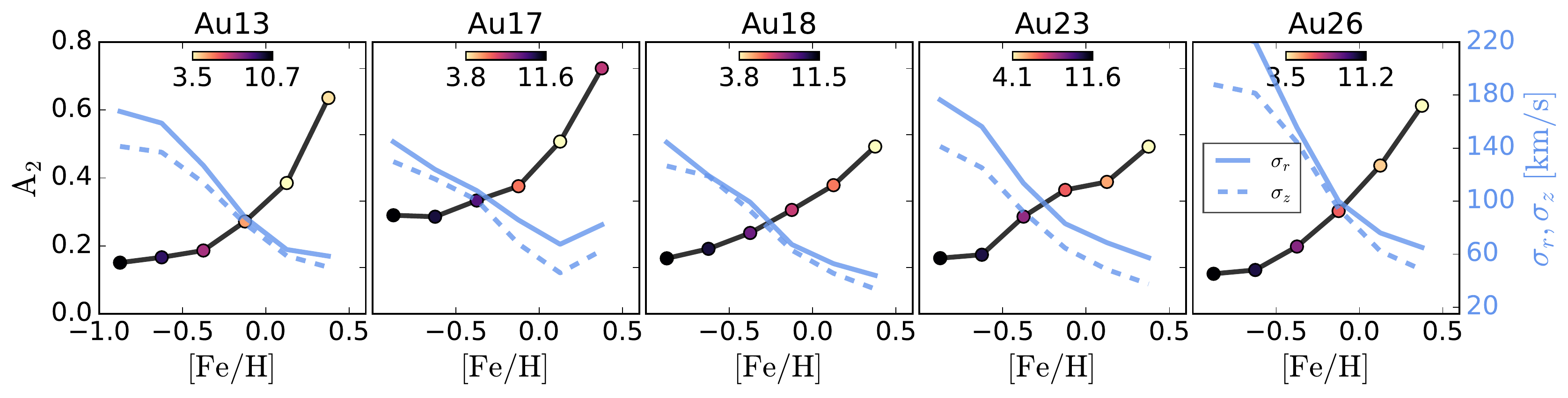}
\caption{The dependence of bar strength on metallicity, velocity dispersion and age: The black curves and left $y$-axis indicate the bar strength $A_2$ for each mono-abundance population. The blue curves and right $y$-axis denote the radial (solid) and vertical (dashed) velocity dispersion of the underlying population, while the colour-coding of the circles indicates their mean age (see colourbar -- in Gyr -- of each panel). Bar strength increases for increasing [Fe/H], decreasing velocity dispersion and younger age of the underlying population.} 
\label{fig:bs_feh_all}
\end{figure*}

\begin{figure}
\centering
\includegraphics[width=0.47\textwidth]{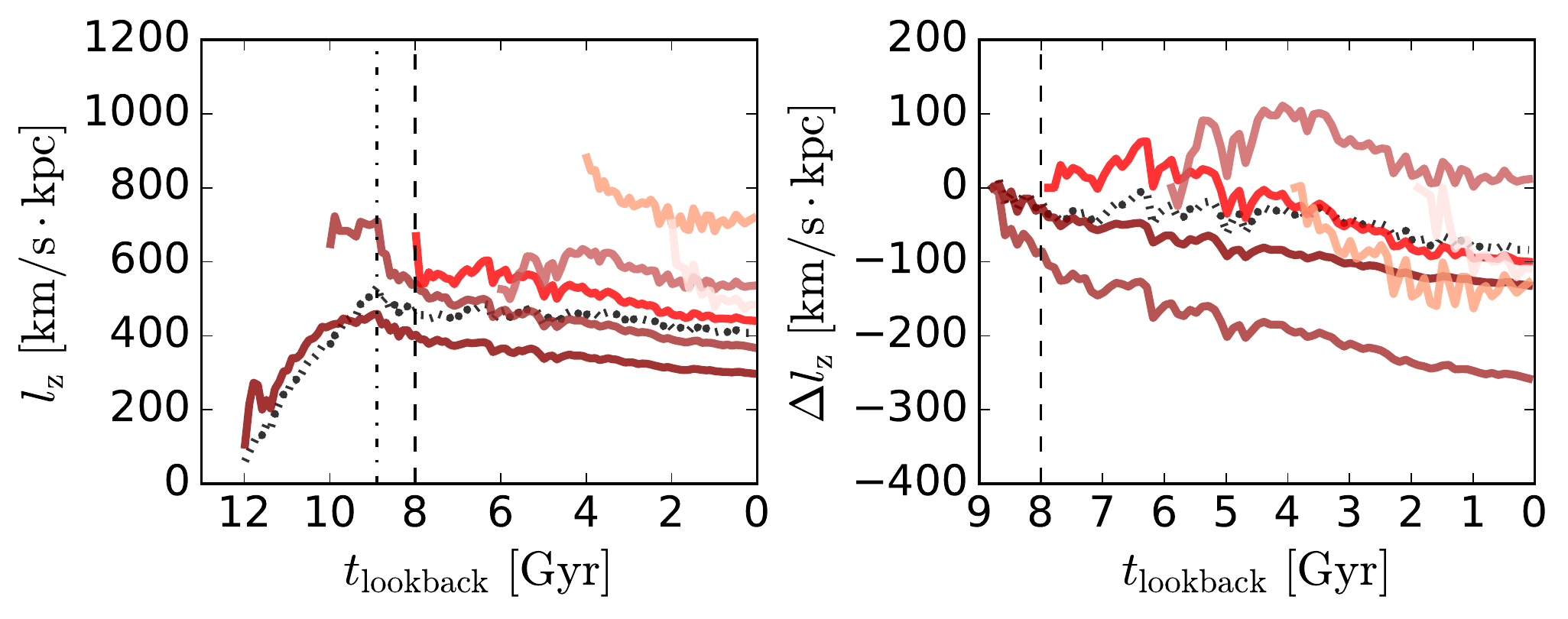}
\includegraphics[width=0.47\textwidth]{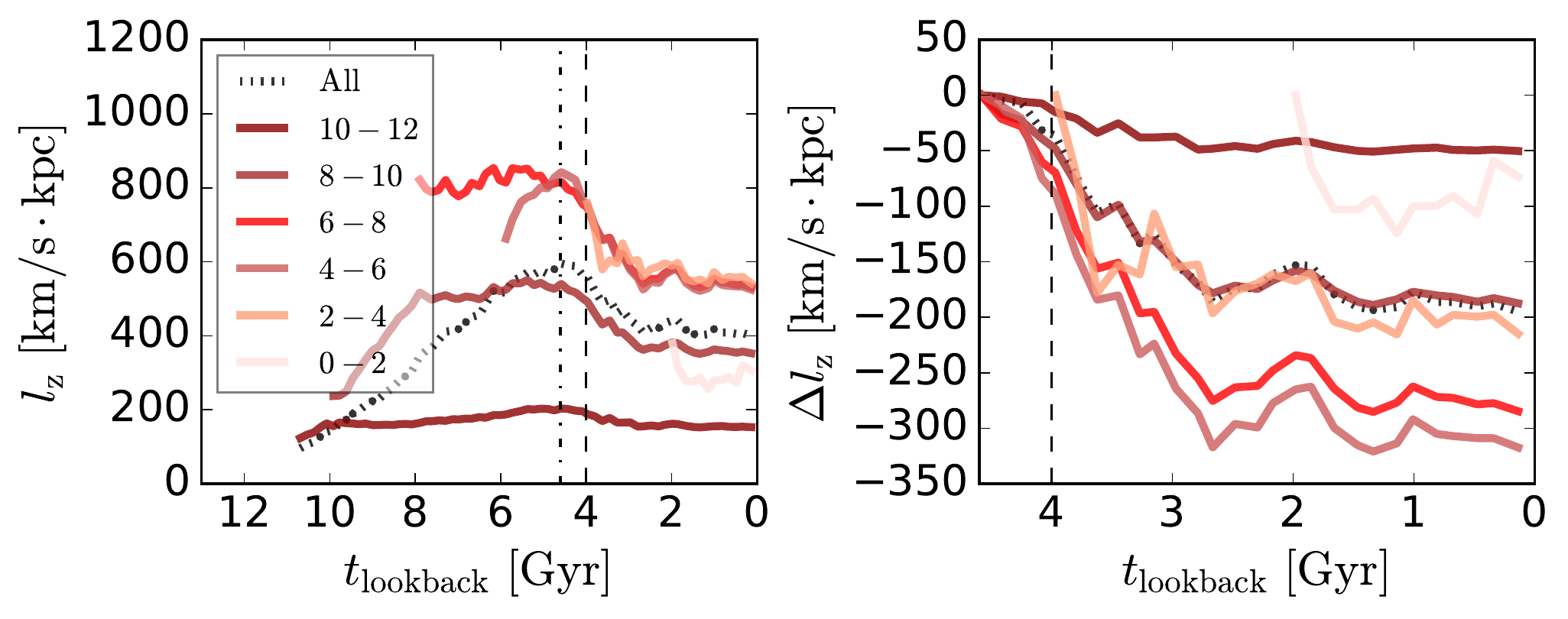}
\caption{\emph{Left:} Evolution of specific angular momentum inside the bar region for mono-age populations (denoted in Gyrs) in Au18 (first row) and Au23 (second row) inside the bar region. The vertical dashed line marks time of bar formation. The vertical dot-dashed line marks onset of bar instability. \emph{Right:} Change in specific angular momentum.} 
\label{fig:angmom2318}
\end{figure}

In this section we examine the chemo-morphological properties of stellar populations in the bars and b/p bulges of the galaxies in our sample.
In Figure \ref{fig:xy_faceon_edgeon} we show mass-weighted face-on surface density maps for different age and metallicity bins for our fiducial model Au18. In the columns of the top row we show the morphology of different mono-age populations where age is indicated in the top left corner of each panel, and in the rightmost panel we show the face-on surface density of the accreted component in the galaxy. In the second row we show the face-on surface density as a function of metallicity, where the metallicity intervals are indicated in the bottom left corner of each panel. The younger and more metal-rich populations have a more elongated, bar-like morphology than the older populations which are on average rounder -- however even these oldest populations show signs of bar-like morphology (see also Figures \ref{fig:xy_faceon_ages_all}-\ref{fig:xy_faceon_feh_all} for the face-on and edge-on surface density maps as a function of age and metallicity for all haloes in our sample). This behaviour has also been found in Made-to-Measure chemodynamical models of the Milky Way bar and bulge (see \citealt{Portailetal2017b}) suggesting that the Milky Way has a similar chemo-morphological dependency in the bar region.

In Figure \ref{fig:bs_feh_all} we quantify this by calculating the bar strength $A_2$ as a function of metallicity for Au18, and the other four galaxies in our sample. In all the haloes in our sample, the bar strength increases as a function of the metallicity of the stellar populations. This behaviour is a consequence of the kinematic properties of the underlying stellar population \citep{Fragkoudietal2017b}, as shown by the blue lines in the same Figure, which denote the in-plane (solid) and vertical (dashed) velocity dispersion, $\sigma_r$ and $\sigma_z$. The coloured circles denote the mean age of each mono-abundance populations, as shown by the colourbar in each panel where the minimum and maximum age are denoted in Gyrs. To calculate the velocity dispersion and age of each population we select stars from the disc in an annulus outside the bar region (so that the velocity dispersion of the stars is minimally affected by the bar)\footnote{By selecting stars at larger radii we bias the mean age towards younger ages, however we are mainly interested in the trend by which younger populations are colder, and therefore have stronger bar-like morphologies.}. We find that the velocity dispersions of the mono-abundance populations decrease for more metal-rich stars while the bar strength increases, signalling the fact that colder, and therefore younger, populations can participate more strongly in the bar instability. We see therefore that there is a relation between the bar (and b/p) morphology and the kinematics of the underlying mono-age or mono-abundance population (for more details see \citealt{Fragkoudietal2017b}; \citealt{Debattistaetal2017} termed this behaviour `kinematic fractionation').

This kinematic differentiation of mono-age populations in the bar and b/p occurs due to the angular momentum that these populations are able to exchange (see also \citealt{Fragkoudietal2017b}). This is further explored in Figure \ref{fig:angmom2318} for our fiducial model, Au18 (top row) and Au23 (bottom row), where we show the specific angular momentum ($l_z$) evolution of mono-age populations in the galaxy inside the bar region (i.e. $R<6\,\rm kpc$). In the left panels the dot-dashed line marks the beginning of the bar instability phase and in all panels the dashed line marks the time at which a strong bar has formed (here we define a strong bar as $A_2=0.3$). 

In the top left panel of the Figure we see that populations are born with progressively more angular momentum, until the bar forms, as denoted by the dot-dashed line. The second column shows the change in specific angular momentum for each population from the onset of the bar instability (or from the time of birth for those which are born after the bar forms). We see that the populations born before the bar lose the most angular momentum (which is redistributed to the outer disc and halo). Of these, the oldest population (10-12\,Gyr; dark brown curve) loses less angular momentum than younger populations (8-10\,Gyr; light brown curve) which are formed just before the bar forms. The same behaviour can be seen for halo Au23 in the bottom panels of Figure \ref{fig:angmom2318}. This occurs because the older populations are hotter and therefore lose less angular momentum than the colder populations which can get trapped on more elongated bar-like orbits (see \citealt{Fragkoudietal2017b}). 
On the other hand, the populations born after the bar forms have an already decreased specific angular momentum, compared to that which they would have were the bar not present. This can be verified by examining the specific angular momentum of mono-age populations in Au23, where the bar forms at a later time as compared to Au18 ($t_{\rm lookback}$=4\,Gyr vs 8\,Gyrs). We see that the specific angular momentum of younger populations increase until the bar forms, as stars are forming on more settled circular orbits. However, once the bar forms, stars born inside the bar region are born on more elongated orbits, thus with less angular momentum to begin with. 

We show the edge-on surface density distributions in the top panels of Figure \ref{fig:morph_edgeon_abund}  in three metallicity bins  -- from top to bottom, [Fe/H]$>$0, -0.5<[Fe/H]<0 and -1<[Fe/H]<-0.5 respectively -- for the galaxies in our sample. As in the case of the face-on projection and the bar, the b/p morphology is more pronounced for more metal-rich populations. However, the morphology of the most metal-poor component (i.e. whether a flattened or spheroidal distribution) is different for each of the five haloes. In Sections \ref{sec:kin2} and \ref{sec:SFH} we will explore how the morphology of the metal-poor population depends on its kinematic properties and on the galaxy's assembly history.


\begin{figure*}
\centering
\includegraphics[width=0.85\textwidth]{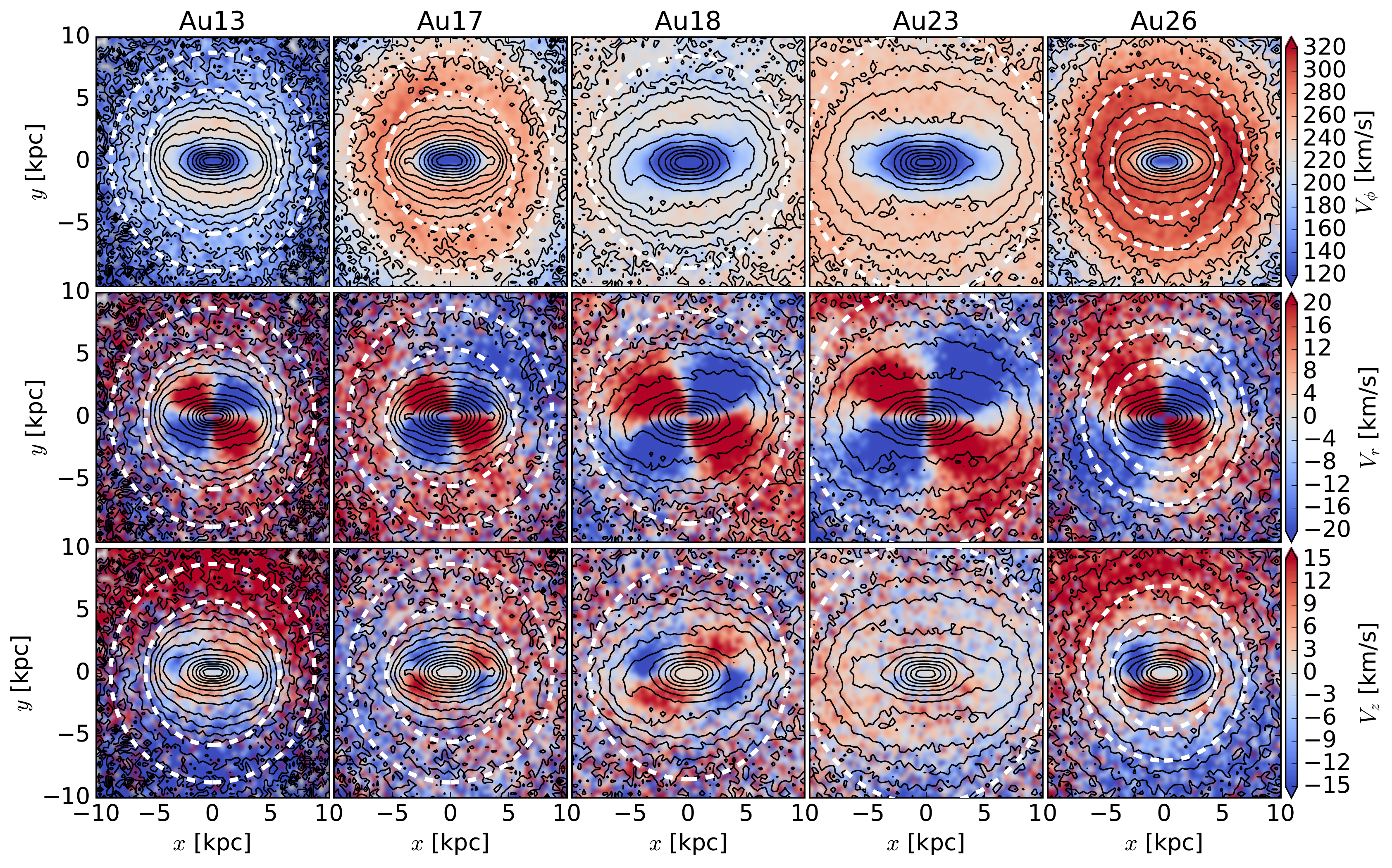}
\caption{Global kinematics in the face-on projection of bars and b/p bulges: Mass-weighted tangential velocity $V_{\phi}$ (first row), radial velocity $V_r$ (second row) and vertical velocity $V_z$ (third row) for stars with $|z|<0.5\,\rm kpc$. In all panels we denote the corotation and OLR radius (when they fall within 10\,kpc) by the inner and outer dashed circles. We see a clear kinematic signature for the asymmetric b/p's (Au17, Au18 and Au26) as a butterfly pattern in $V_z$ (see Appendix \ref{sec:Appendixvels} for more details).} 
\label{fig:xyall}
\end{figure*}

\begin{figure*}
\centering
\includegraphics[width=0.86\textwidth]{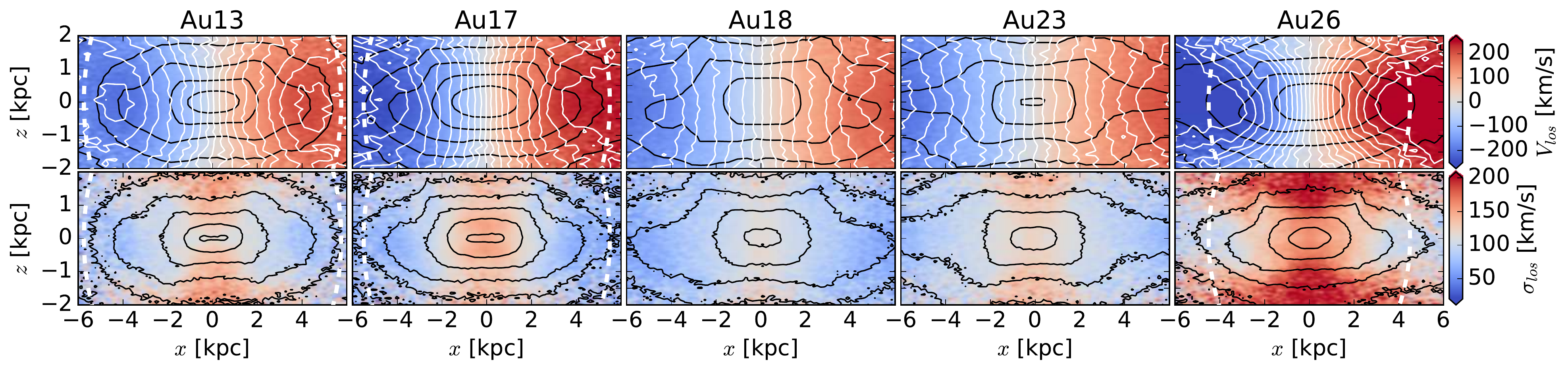}
\caption{Global kinematics in the edge-on projection of bars and b/p bulges: \emph{First row:} Mass-weighted line of sight velocity where iso-velocity contours are denoted with white lines with a spacing of $25\, {\rm km\,s^{-1}}$. We see that all models show cylindrical rotation. \emph{Second row:} Mass-weighted line of sight velocity dispersion. We see that the velocity dispersion profile has an X-shape. In all panels the vertical dashed line indicates the CR radius of the model (if it is inside 6\,kpc), the bar is along the $x$-axis and the black lines show iso-density contours.} 
\label{fig:xzvelsall}
\end{figure*}

\begin{figure*}
\centering
\includegraphics[width=0.95\textwidth]{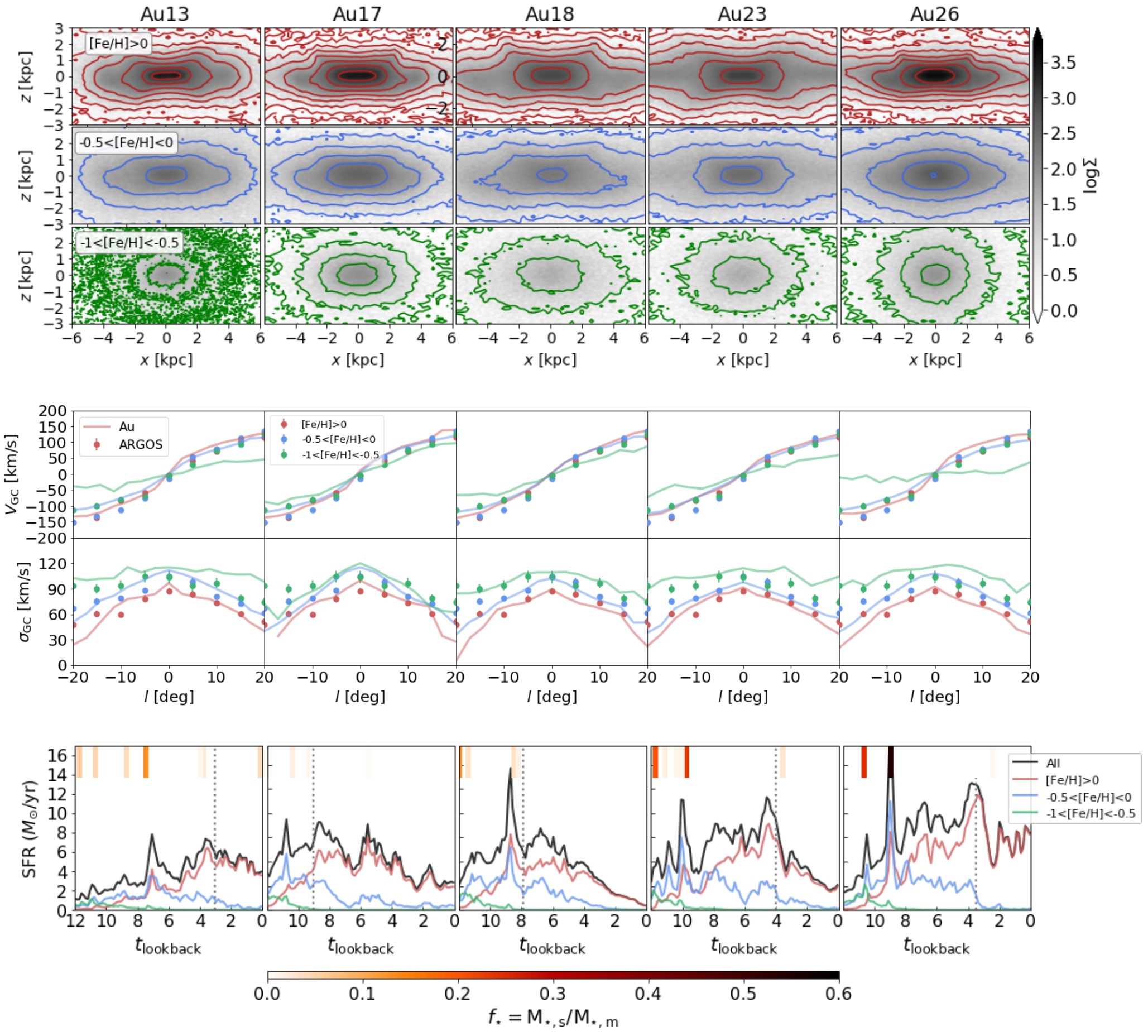}
\caption{{\textbf{The relation between chemodynamical properties and formation history:}} \emph{Top panels:} Edge-on morphology of b/p's in three metallicity bins: From top to bottom, [Fe/H]>0, -0.5 <[Fe/H]<0 and -1 <[Fe/H]<-0.5.
\emph{Middle panels:} Line of sight velocities (top) and velocity dispersion (bottom) for stars in the three aforementioned metallicity ranges as a function of longitude. We compare these to the kinematic properties of the Milky Way bulge (circles) from \citet{Nessetal2013b}.
\emph{Bottom panels:}
Star formation history of the disc for the five galaxies in our sample (black line), and separated in the three different metallicity bins as indicated by the coloured lines in the legend. The vertical coloured lines at the top of the panels indicate the merger time of all galaxies with $M_{\star,s} > 10^7\rm M_{\odot}$; the colour of the lines indicates the stellar mass ratio of the merger, $f_{\star} = M_{\star,s}/M_{\star,m}$ (see the colourbar below the figure). The vertical black dotted line marks the formation time of the bar. The galaxies in which the metal-poor component has a thick-disc-like morphology (e.g. Au17 and Au18) are also the ones in which these stars rotate almost as fast as the metal-rich population. These galaxies are also the ones in which the merger history of the galaxy is very quiescent, indicating the intimate relation between the chemodynamical properties of the bulge and formation history.
} 
\label{fig:morph_edgeon_abund}
\end{figure*}

\section{Kinematic properties of stellar populations in bars and b/p bulges} 
\label{sec:chemokinem}

In this section we examine the global kinematic properties of the bars and b/p bulges in Auriga (Section \ref{sec:kin1}), and then explore the kinematic properties of mono-abundance populations inside the bar-b/p region, comparing them to the properties of the Milky Way bulge (Section \ref{sec:kin2}). 

\subsection{Global Kinematic properties}
\label{sec:kin1}
In Figure \ref{fig:xyall} we show face-on kinematic maps (mass-weighted mean $V_{\phi}$, $V_r$ and $V_z$) of the Auriga haloes in our sample; we focus on the kinematics close to the plane by taking stars within $|z|<0.5\, \rm kpc$. The corotation and OLR radii are marked with the inner and outer dashed circles respectively, while the black lines are iso-density contours of the face-on surface density distribution.
In the top row of Figure \ref{fig:xyall} we see that all galaxies show an elongated shape of low $V_{\phi}$, which follows the shape of the bar, due to the slower rotation of stars on bar-like orbits. In some haloes (e.g. Au18 and Au23) we also see an X-shape in the $V_{\phi}$ pattern, indicating the X-shaped morphology of orbits in the peanut region. 
The face-on distribution of $V_r$ shows a butterfly pattern, characteristic of barred galaxies, i.e. of inward and outward moving velocities in the bar region. 
When examining the mean vertical velocity, $V_z$, we see that there is a butterfly pattern in some of the haloes, in particular in Au17, Au18 and Au26. Upon closer examination of the edge-on morphology of these haloes (top row of Figures \ref{fig:morph_edgeon_abund} \& \ref{fig:xy_edgeon_ages_all}) we see that their boxy/peanuts are asymmetric, i.e. the peanut is currently undergoing a buckling instability. This signature of buckling bars in velocity was first explored in an isolated N-body simulation of a Milky Way-like galaxy in \citet{Lokas2019}, and we confirm this signature here using the Auriga cosmological simulations, as well as verifying the signature with an isolated disc galaxy simulation in Appendix \ref{sec:Appendixvels}. Interestingly, Au17 and Au18 which are currently undergoing a buckling phase, have bars which formed at $z>1$ and have had a b/p since 8 and 5\,Gyrs ago (see the thick dashed line in the bottom panel of Figure \ref{fig:rgball}). This indicates that they are undergoing a renewed buckling instability at $z=0$, which has been shown to occur for strong bars (e.g. \citealt{MartinezValpuestaetal2006}). This butterfly pattern in $V_z$, which is present for asymmetric b/p's, can therefore be used to identify buckling bars in almost face-on projections.

In Figure \ref{fig:xzvelsall} we show edge-on kinematic maps of the galaxies in our sample (where the bar is aligned with the $x$-axis), focusing on the inner region, i.e. on the boxy/peanut bulges. To reduce contamination from outer disc particles, we select stars within $R<6\,\rm kpc$ of the galactic centre. In the cases where corotation falls inside the panel, it is marked with a curved white dashed line. In the first row we show the mass-weighted line-of-sight velocity $V_{\rm los}$ for the galaxies in our sample, with white contours denoting iso-velocity lines with spacing of 25\,km/s. As expected for boxy/peanut bulges, these galaxies exhibit cylindrical rotation inside the boxy/peanut bulge region, i.e. the line of sight velocity is independent of height above the plane. In \citetalias{Gargiuloetal2019} we also showed that, overall, the bulges in Auriga exhibit rather large amounts of rotation. In the second row of Figure \ref{fig:xyall} we show the mass-weighted line of sight velocity dispersion, $\sigma_{\rm los}$, which for all haloes has a distinctive X-shape, with low velocity dispersion tracing the tips of the peanut.

\subsection{Chemo-kinematic properties}
\label{sec:kin2}


We now explore the kinematic properties of stars in the bar-b/p of the Auriga galaxies as a function of metallicity, and compare these to the chemo-kinematic properties of the Milky Way bulge. 
In the middle panels of Figure \ref{fig:morph_edgeon_abund} we show the line of sight velocity, $V_{\rm GC}$, and velocity dispersion, $\sigma_{\rm GC}$, as a function of longitude for stars in the b/p bulges of the Auriga galaxies (solid lines). We separate the stars into three metallicity bins --  [Fe/H]>0 (red), -0.5<[Fe/H]<0 (blue) and -1<[Fe/H]<-0.5 (green) and compare these to the velocity and velocity dispersion of stars with corresponding metallicities in the Milky Way bulge (circles), using data from the ARGOS survey \citep{Nessetal2013b}. In order to compare the models to the Milky Way, we rescale the masses of the Auriga galaxies to match the stellar mass of the Milky Way, and rotate the bar to have an angle of 30 degrees with respect to the galactocentric line of sight, placing the observer at 8.3\,kpc from the centre of the galaxy \citep{BlandHawthornGerhard2016}. For this plot we select stars close to the plane ($b<1\,\rm degrees$ for the model and $b=5\,\rm degrees$ for the ARGOS data).

In the Milky Way bulge the line-of-sight velocity $V_{\rm GC}$ of stars is comparable for the three different metallicity bins, i.e. even the most metal-poor stellar population with -1<[Fe/H]<-0.5 has significant rotation, similar to the metal-rich and intermediate populations, even though it is a hotter component with higher velocity dispersion. For the models explored here, this behaviour approximately holds for Au17 and Au18, i.e. all three metallicity components have similar rotation. On the other hand, we see that the most metal-poor component in models such as Au26 has little net rotation, differing significantly from the rotation of the metal-rich and intermediate components. We also note the correlation between morphology and rotation for the metal-poor component, i.e. Au17 and Au18 which have highly rotating metal-poor populations, also have flattened, thick disc-like morphologies for the metal-poor populations, while Au26 whose metal-poor component is hardly rotating has a spheroidal morphology. As we discuss below, in Section \ref{sec:SFH}, the merger history of these galaxies is imprinted on the morphology and kinematics of the metal-poor stellar populations. 

\begin{figure}
\centering
\includegraphics[width=0.49\textwidth]{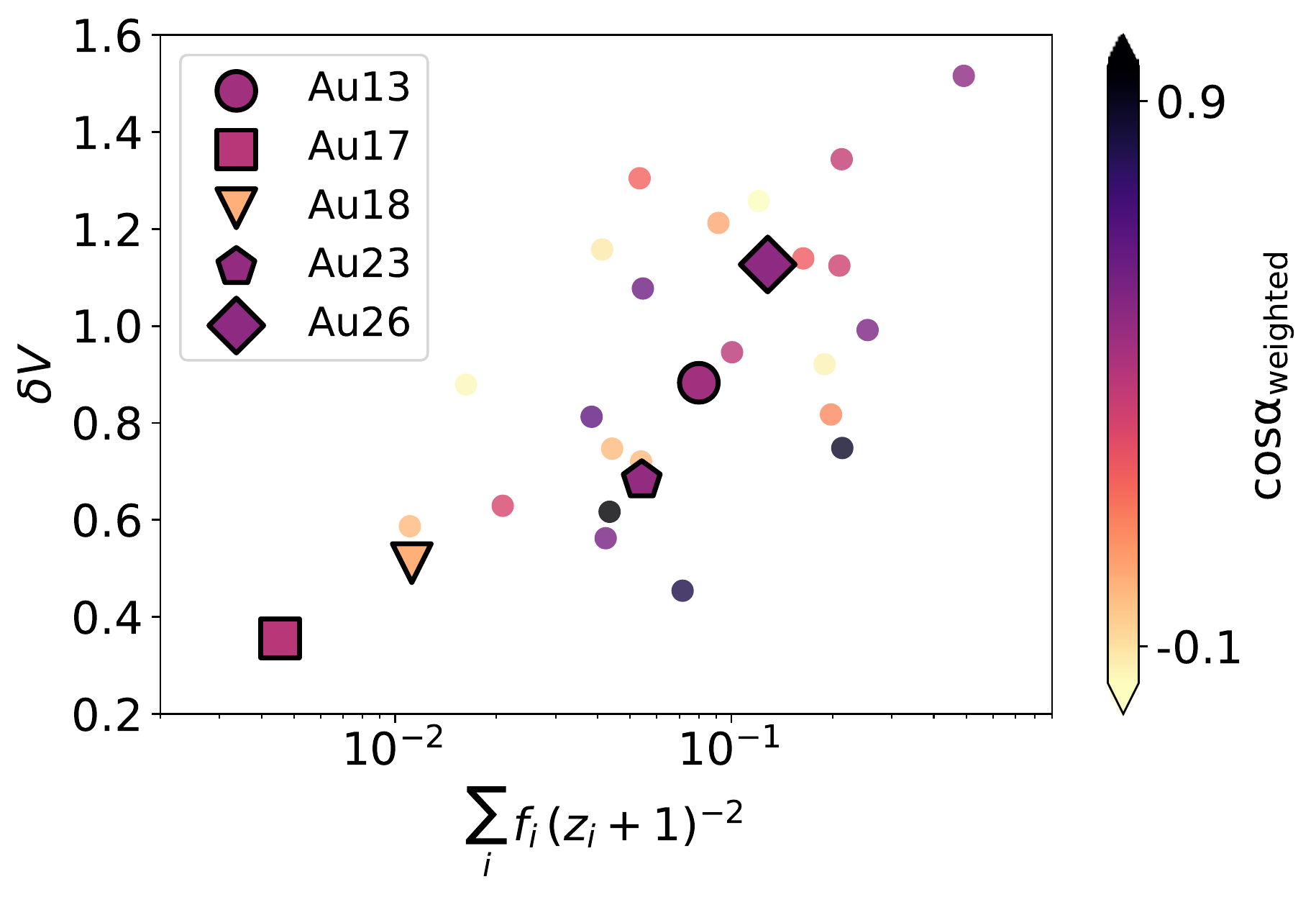}
\caption{{\textbf{$\delta V$ vs merger impact:}} The normalised difference in velocity between the metal-rich, intermediate and metal-poor populations at the effective radius ($\delta V$) of all Auriga galaxies versus the impact of mergers, given by the sum over all mergers of the stellar mass ratio normalised by the merger redshift (see text for more details). The colour-coding shows the sum of the dot product of the angular momentum of the main disc and that of the orbital angular momentum of the merging subhalo, weighted by the merger impact, indicating how prograde or retrograde all the mergers in a given galaxy are. Disc galaxies with higher impact from mergers will have larger $\delta V$, as discussed in Figure \ref{fig:morph_edgeon_abund}; Haloes Au17 and Au18 which are most similar to the Milky Way in terms of bulge chemodynamics have the lowest merger impact and overall number of mergers with stellar mass above $10^7\rm M_{\odot}$. Some of the scatter in the relation is due to the orbital configuration of mergers: galaxies with cos$\alpha$ $\sim 1$ have had mostly prograde mergers which lead to lower $\delta V$ for a higher merger impact.} 
\label{fig:formrelation}
\end{figure}

\section{Formation histories of barred-b/p galaxies}
\label{sec:SFH}

\begin{figure}
\centering
\includegraphics[width=0.49\textwidth]{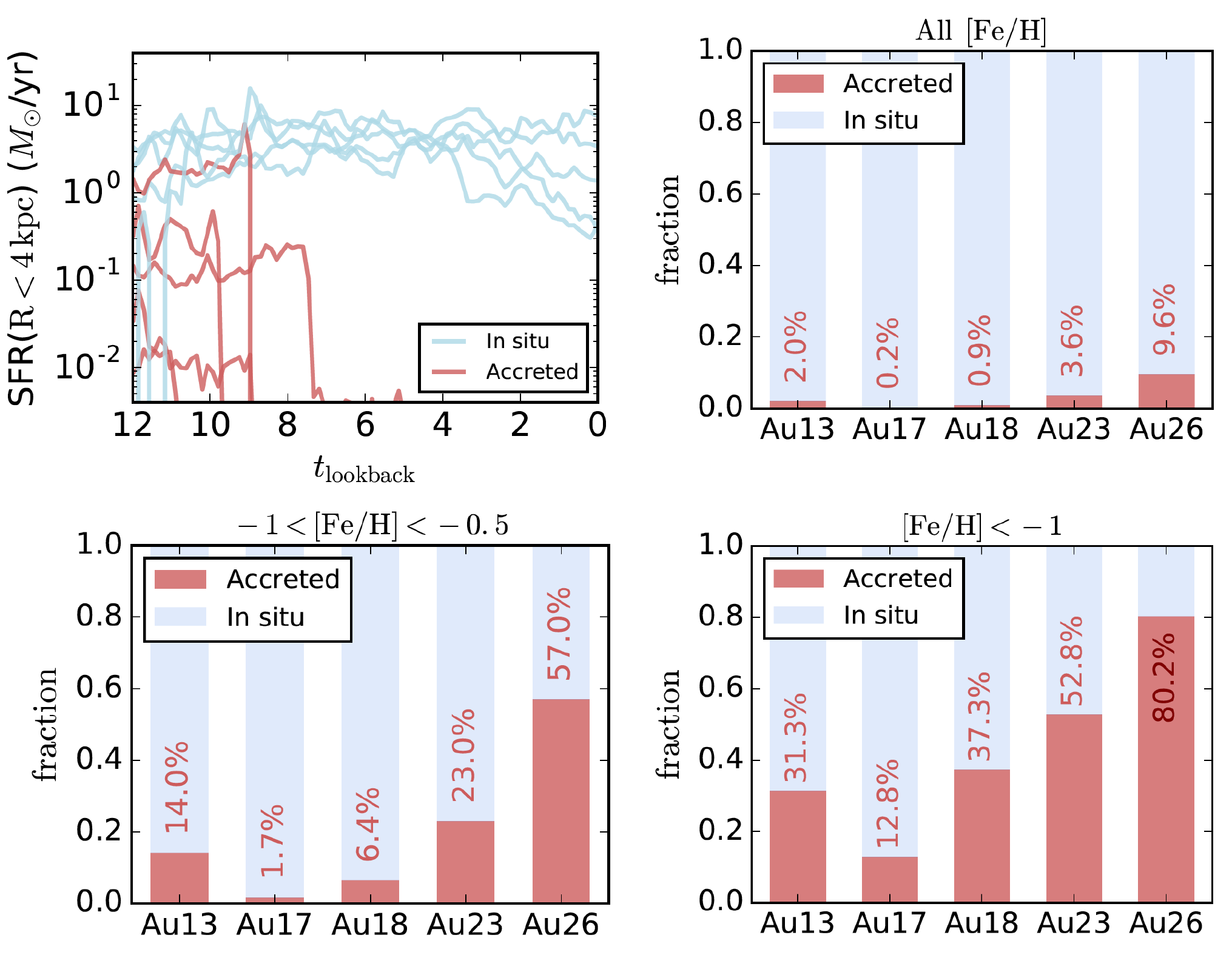}
\caption{Comparison of in-situ vs accreted population in the boxy/peanut bulges of our sample. \emph{Top left:} Star formation histories for the in-situ (blue) and accreted (red) populations in the haloes in our sample. We see that for all haloes the accreted material in the b/p region is older than 7\,Gyrs. \emph{Top right:} Fraction of accreted vs in-situ stars inside the b/p bulge region for the five haloes for stars of all metallicities. \emph{Bottom panels:} Fraction of in-situ vs accreted stars for stars with metallicities -1$<$[Fe/H]$<$-0.5 (left) and [Fe/H]<-1 (right).} 
\label{fig:fr_accr}
\end{figure}

\begin{figure}
\centering
\includegraphics[width=0.35\textwidth]{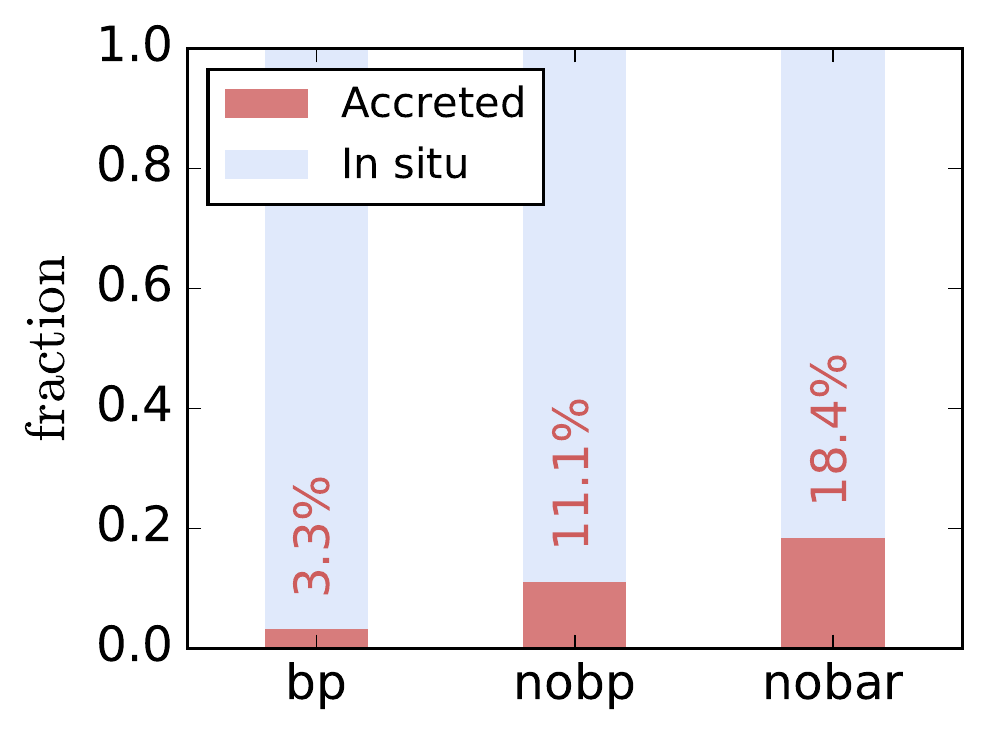}
\caption{Average fraction of accreted vs in-situ stars in the central regions ($r<4\,\rm kpc$, $|z|<2\,\rm kpc$) of galaxies in Auriga, separating into those in this study, i.e. with a boxy/peanut (bp), barred Auriga galaxies without a prominent b/p (nobp) and Auriga galaxies without a bar (nobar). We see that galaxies which form a b/p have on average little accreted material in their central region, with the ex-situ fraction being of the order of a few percent.} 
\label{fig:fr_withwithoutbp}
\end{figure}

We explore the link between the formation history of galaxies in our sample and the chemo-morphological and chemo-kinematic properties of stellar populations in their bars and b/p bulges (for an exploration of the formation histories of \emph{all} bulges in Auriga see \citetalias{Gargiuloetal2019}).
In the bottom panels of Figure \ref{fig:morph_edgeon_abund} we show the star formation rate (SFR) in the disc (solid black curve) of the five galaxies in our sample, i.e. within $R<20\,\rm kpc$ and $|z|<2\,\rm kpc$. The red, blue and green curves indicate the SFR of the metal-rich ([Fe/H]>0), intermediate (-0.5<[Fe/H]<0) and metal-poor (-1<[Fe/H]<-0.5) populations respectively. In all panels the vertical black dotted line marks the formation time of the bar. The upper thick vertical coloured lines mark the merger times of all satellites with stellar mass $M_{\star,s}>10^7 \rm M_{\odot}$; the lines are coloured according to the stellar mass ratio of the merger $f_{\star}=M_{\star ,s}/M_{\star , m}$, where $M_{\star ,s}$ and $M_{\star,m}$ are the stellar mass of the satellite and the main galaxy respectively, at the time of satellite infall\footnote{We note that the bars in our sample tend to form soon after a merger, since mergers increase the disc mass, and can also remove angular momentum from the disc thus kick-starting the formation of the bar (we will discuss in more detail the mechanisms responsible for bar formation in Auriga galaxies in Fragkoudi et al. in prep.).}. 

By comparing the middle and bottom panels of Figure \ref{fig:morph_edgeon_abund}, we see that the two haloes that have the most similar kinematic properties to the bulge of the Milky Way -- in terms of the high rotation of their metal-poor component -- i.e. Au17 and Au18, have no major mergers (stellar mass ratios$>$1:4) in the last 12\,Gyrs of their evolution. The most massive mergers these galaxies experience since $z\sim3.5$ are $<$1:20 mergers, which is broadly consistent, although on the low mass end, with respect to recent estimates of a Gaia Sausage/Enceladus-type merger for the Milky Way\footnote{As discussed in \citealt{Monachesietal2019}, we caution that the Auriga haloes are in general more massive than the Milky Way halo, so the limit on the most massive accreted system in the Milky Way will likely be lower. We also point out that, as shown in \citealt{Monachesietal2019},  40\% of the halo mass comes from the Gaia Enceladus-like progenitor, however a total of 14 satellites make up the entire halo of this model.} (e.g. \citealt{Haywoodetal2018,Helmietal2018,Belokurovetal2019,Deasonetal2019}; see also \citealt{Bignoneetal2019}). These Enceladus-like mergers in Au17 and Au18 also result in stars in the inner halo being on radial orbits \citep{Fattahietal2019}. 


On the other hand, Au13, Au23 and Au26 have significant ($f_{\star}>0.1$) mergers occurring in the last 12\,Gyrs. Halo Au26 undergoes a massive 1:2 merger at $t_{\rm lookback}\sim9\, \rm Gyrs$ which creates a dispersion dominated, spheroidal bulge for the metal-poor, old populations (see Figures \ref{fig:morph_edgeon_abund} \& \ref{fig:xy_edgeon_ages_all}). It is worth noting that the metal-poor population was already mostly formed at the time of the merger (see green line in bottom right panel of Figure \ref{fig:morph_edgeon_abund}) and thus the merger disrupts this in-situ, old and metal-poor component, creating a dispersion dominated spheroid, while the new stars born during this merger are also on non-circular orbits (see also \citealt{Grandetal2020} for a discussion on the effects of an Enceladus-like mergers on the disc and halo populations). Haloes Au13 and Au23 also have metal-poor populations with lower net rotation at $z=0$ (first and fourth column of middle panels Figure \ref{fig:morph_edgeon_abund}) as they undergo significant mergers in their recent past -- Au13 undergoes a $f_{\star}=0.15$ merger at $t_{\rm lookback}=7\,\rm Gyrs$, while Au23 undergoes a $f_{\star}=0.25$  merger at $t_{\rm lookback}=10\,\rm Gyrs$. 

We therefore find that there is an upper limit on how massive mergers can be (since $t_{\rm lookback}\sim12\,\rm Gyrs$) while still maintaining a rotationally supported metal-poor component in the inner regions of disc galaxies. This behaviour is summarised in Figure \ref{fig:formrelation}, where we plot the normalised difference in rotation velocity at $z=0$ between the metal-rich, intermediate and metal-poor populations at the effective radius $\delta V = (V_{\rm MR} - V_{\rm INT})/V_{\rm MR} + (V_{\rm MR} - V_{\rm MP})/V_{\rm MR}$, versus the impact of mergers, which we here define as the sum of the mass ratio of mergers (for $M_{\star,s}\geq10^7\rm M_{\odot}$) divided by the redshift of the merger squared, i.e. the impact of mergers is given by $\sum_{\rm i} f_{\rm i}/(z_{\rm i} +1)^2$. There is a trend for the five galaxies we explore in our sample, which are denoted by the large symbols (see the legend), in which the larger the merger impact, the larger the difference in rotation between the stellar populations $\delta V$. We also include in the figure the rest of the Auriga galaxies (small circles) for the entire parent sample of 30 (\citealt{Grandetal2017}; excluding  two Auriga haloes which are undergoing a merger at $z=0$). These follow a similar trend, albeit with considerable scatter. We colour-code the symbols by the sum of the cosines of the angles between the angular momentum vector of the disc of the main galaxy and that of the orbital angular momentum of each merging subhalo, weighted by the impact of each merger; this indicates how prograde, or retrograde the mergers are. We see that some of the scatter in the relation comes from the fact that galaxies which undergo more prograde mergers can have relatively high merger impact while having low $\delta V$, while galaxies undergoing retrograde mergers will have higher $\delta V$ for a smaller merger impact. We note that haloes Au17 and Au18, which are the most Milky Way-like haloes in terms of the overall morphology and chemodynamics of their bulge stellar populations, have the most quiescent merger histories in the entire Auriga sample. This highlights the importance of a quiet merger history for forming b/p bulges which have chemo-kinematic properties similar to the Milky Way bulge.

\subsection{Ex-situ fraction of stars}
We now explore the amount of ex-situ stars in the b/p bulges in Auriga, and how this relates to their chemodynamical properties.
As shown in \citetalias{Gargiuloetal2019}, Auriga bulges have a range of ex-situ fractions, from $<1\%$ to 42\%, with many of the bulges forming mostly in-situ, i.e. 21\% of the Auriga galaxies have less than 1\% of ex-situ fractions.  
In the top left panel of Figure \ref{fig:fr_accr} we show the star formation histories in the b/p bulge region (i.e. $R<4\,\rm kpc$ and $|z|<2\,\rm kpc$) for the accreted (red) and in-situ (blue) populations, and in the top right panel we list the fraction of accreted to in-situ stars inside the b/p's. Most stars present in the central regions of our sample of galaxies are formed in-situ, with almost all ex-situ material being accreted at early times, before $z\sim1$ (see also \citealt{Bucketal2019} who similarly found low ex-situ fractions in the central regions of their cosmological model). Therefore the accreted stars in the b/p bulges of our models are subdominant in all haloes, less than 1\% for Au17 and Au18, of the order of a few percent for Au13 and Au23, with the highest fraction of ex-situ stars being found in Au26 which has 9\% of ex-situ stars. 

We therefore find that the fraction of ex-situ stars is also linked to the kinematics of the metal-poor populations of the bulge; Au17 and Au18 (which have the smallest ex-situ fractions) have metal-poor populations which rotate fast, contrary to Au13, Au23 and Au26, which have higher ex-situ fractions and more slowly rotating metal-poor populations.
As expected of course, the fraction of ex-situ stars increases as we consider lower metallicity ranges, as shown in the bottom left panel of Figure \ref{fig:fr_accr}. Haloes Au17 and Au18 -- which have similar kinematics to the MW -- have only a few percent  (1.7 and 6.4\% respectively) of ex-situ stars even in the metal-poor population of -1<[Fe/H]<-0.5, while Au26 has almost 60\% of the metal-poor population in the ex-situ component (and see also \citealt{Monachesietal2019} for the fraction of accreted material in the haloes of these galaxies). If we consider stars with [Fe/H]$<$-1 (bottom right panel of Figure \ref{fig:fr_accr}), the fraction of accreted stars increases substantially, with a maximum of 80\% (for Au26) while it can still be quite low, for example 13\% for Au17, which has a very quiescent merger history, and $\sim$38\% for our fiducial Milky Way model, Au18. Therefore, in order to detect the ex-situ population of stars in the MW b/p bulge we will likely have to probe the extremely metal-poor tail of the inner regions (see e.g. \citealt{Starkenburgetal2017,Arentsenetal2019} and Figure \ref{fig:vlos_sigmalos_all_append} where we show the kinematics of the [Fe/H]<-1 population in Auriga).

We also find that the low fraction of ex-situ stars is a generic property of b/p bulges, compared to other bulges\footnote{For a discussion of all bulges in Auriga see also \citetalias{Gargiuloetal2019}.}. This is shown in Figure \ref{fig:fr_withwithoutbp}, where we consider the fraction of ex-situ stars in this sample (bp), compared to haloes in Auriga which have bars but do not form b/ps (nobp), and those that do not form bars at all by $z=0$ (nobar). As previously, we consider as inner regions those inside $r<4\,\rm kpc$ and $|z|<2\,\rm kpc$. We find that haloes with bars that form b/p's tend to have the smallest fraction of ex-situ stars, compared to those which do not form b/p's, while haloes without bars have the highest fraction of ex-situ stars in their bulge region. This is due to the intimate connection between the presence of a dispersion dominated component in the central regions of disc galaxies, and the formation of bars and b/p's (e.g. \citealt{Athanassoula2005}) and highlights the importance of relatively quiescent merger histories for the formation of b/p bulges.

\section{effect of the bar on the disc}
\label{sec:bareffectdisc}

In this Section we explore the effects of bars on their host discs, focusing in Section \ref{sec:effectring} on the effects of the bar on the inner disc, via the formation of so-called `inner rings', and in Section \ref{sec:phasespace} on their effects on the outer disc, via the formation of ridges in phase space.

\subsection{Effect of the bar on the inner disc: Formation of inner rings}
\label{sec:effectring}

\begin{figure*}
\centering
\includegraphics[width=0.95\textwidth]{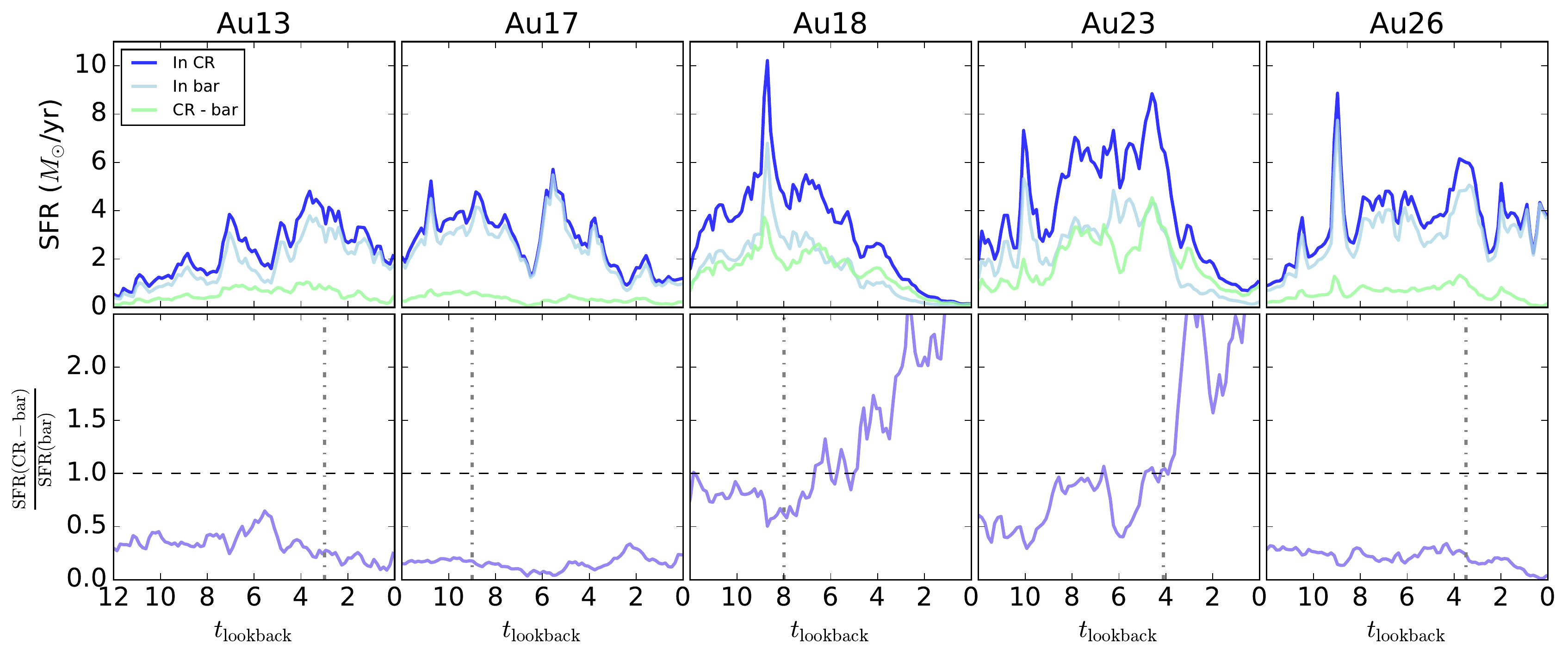}
\caption{\emph{Top row:} Star formation history for the five haloes in our sample, inside the corotation radius (blue), inside the bar (light blue) and the difference between the two (i.e. CR-bar; light green). When there is no inner ring (i.e. Au13, Au17 and Au26), the (CR-bar) region corresponds to the so-called star formation desert, while in the presence of an inner ring it will correspond to the star formation inside the inner ring. \emph{Bottom row:} The SFR of the (CR-bar) region over the SFR of the bar region. In all panels the vertical dot-dashed line corresponds to the formation time of the bar. We see that in the cases without an inner ring the ratio stays below 1, while in the case where an inner ring forms, the ratio increases once the bar and inner ring are in place. For Au18 this corresponds to $t_{\rm lb}\sim7\, \rm Gyr$ and for Au23 to $t_{\rm lb}\sim9\, \rm Gyr$. This method could be used to age date the formation of the bar in the presence of an inner ring.}
\label{fig:sfhring}
\end{figure*}

\begin{figure*}
\centering
\includegraphics[height=0.2\textwidth]{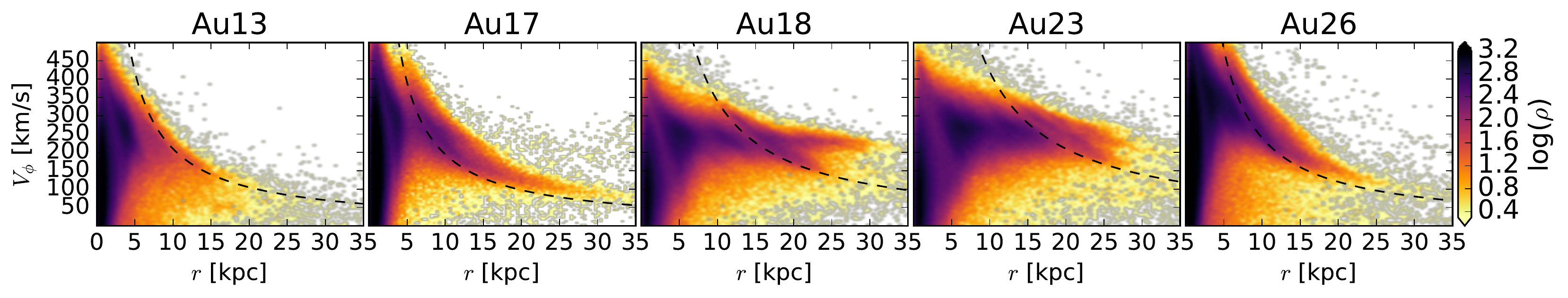}
\caption{Logarithmic density in the $V_{\phi}-r$ plane for all haloes in the study. The dashed line indicates the constant angular momentum of a circular orbit at the OLR radius. We see that in all cases where the disc extends beyond the OLR (i.e. all apart from Au13) the longest and most prominent ridge corresponds to the OLR.} 
\label{fig:vphir_all} 
\end{figure*}

\begin{figure*}
\centering
\includegraphics[width=0.75\textwidth]{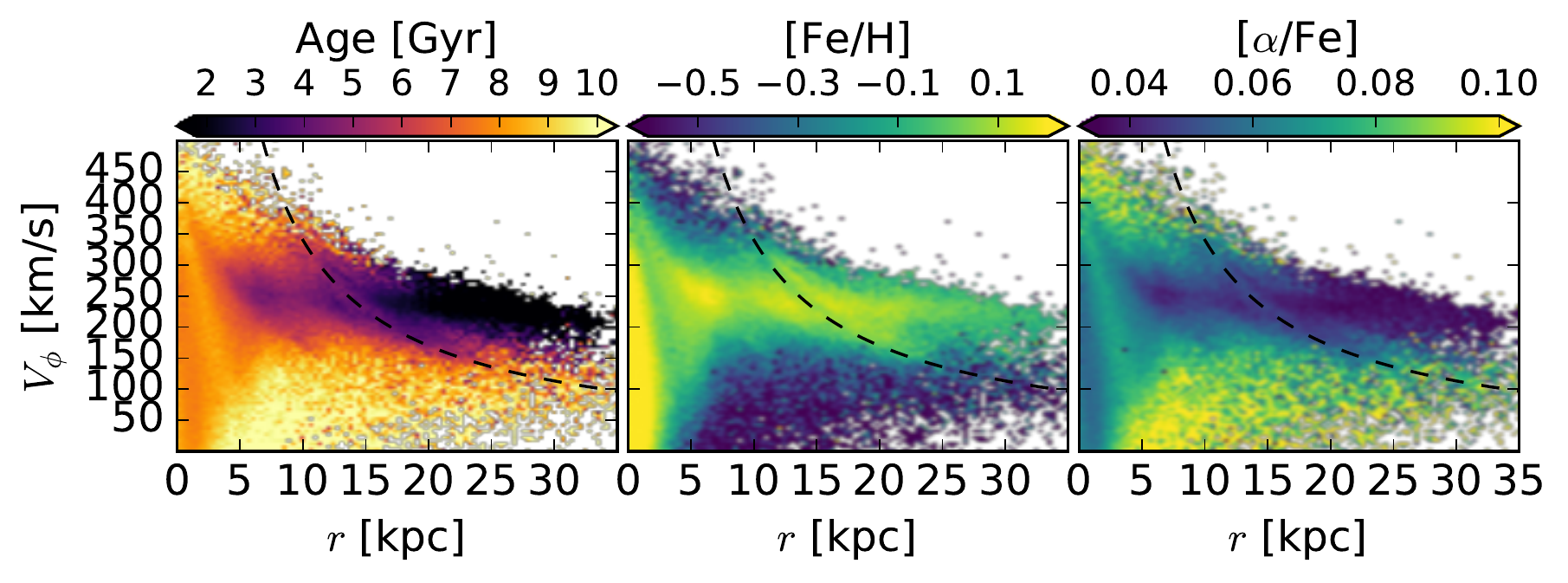}
\caption{The $V_{\phi}-r$ plane for fiducial model Au18 with mean mass-weighted age (left), metallicity (middle) and $\alpha$-abundance (right) colour-coded. The dashed line indicates the angular momentum of a circular orbit at the OLR radius. We see that the ridges have younger, more metal-rich and more $\alpha$-poor stars on average than surrounding regions of phase-space.} 
\label{fig:vphir_au18_agefe} 
\end{figure*}

By examining the age and abundance maps of the galaxies in our sample in Section \ref{sec:ageabund}, we see that in Au18 and Au23 the bar affects the age, metallicity and $\alpha$-abundance distribution of the disc with the formation of a prominent inner ring. 
The formation mechanism of inner rings is still under debate, however they have classically been associated with the locations of resonances in barred galaxies (e.g. \citealt{Schwarz1981,ButaCombes1996}). A more recent theory which was proposed to explain the presence and morphology of inner and outer rings in barred galaxies is the invariant manifold theory (\citealt{RomeroGomezetal2006,AthanassoulaRomeroGomez2009}). Invariant manifolds can be thought of as `tubes' which emanate from the ends of the bar, and which guide stars and gas along particular orbits, and which are able to transport material from outside to inside corotation and vice versa.

Such metal-rich and star-forming inner rings could partially explain the recently reported metal-rich ``inner disc'' in the Milky Way which was reported using APOGEE and Gaia DR2 data \citep{Bovyetal2019}. Firstly, the metal-poor inner regions of the bar and the positive metallicity gradient with radius reported in \citet{Bovyetal2019}, can be partially explained by the interplay between the metal-poor thick disc and the metal-rich thin disc of the Milky Way: in the innermost regions the metal-poor thick disc dominates, as it has a shorter scale-length, while at larger radii the metal-rich thin disc starts to dominate.  This was explored in the model presented in \citet{Fragkoudietal2018}, where it was shown (see Figure 13) that the face-on metallicity maps of a composite thin+thick disc model produce a low metallicity innermost region, with metallicity increasing as a function of radius (note that this model was constructed to reproduce observations of the bulge of the Milky Way, and so serves as a verified prediction for the relation shown in \citealt{Bovyetal2019}). However, in the thin+thick disc model of \citet{Fragkoudietal2018}, the metallicity of the inner disc does not reach the highest values (0.2\,dex) found in \citet{Bovyetal2019} around the bar. Reaching these high metallicities could however be naturally achieved with the addition of a star forming inner ring due to the bar. As we show below, in these inner rings, ongoing star formation can help them reach higher metallicities. Indeed our galaxy is thought to host a gaseous inner ring as observed in both HI and CO, called the near and far 3\,kpc arms \citep{vanWoerdenetal1957,DameThaddeus2008}. 

In Figure \ref{fig:sfhring} we show the star formation histories inside corotation for the five haloes in different regions: the blue line indicates the star formation rate inside the entire corotation region, excluding the inner kpc where the nuclear disc (or pseudo-bulge) is. In the light blue line we show the star formation rate in the bar (excluding the inner 1\,kpc) and in light green we show the star formation rate inside corotation minus that inside the bar. In the absence of an inner ring this region (CR - bar) will correspond to the so-called star formation desert (see e.g. \citealt{Jamesetal2009,DonohoeKeyesetal2019}), while in the presence of an inner ring this region will correspond to star formation inside the ring.

The fact that star formation in the inner ring (which is inside corotation) continues, while star formation in the bar is quenched (see the green lines in the top panels of Figure \ref{fig:sfhring}, for Au18 and Au23), suggests that gas inside corotation is constantly being replenished. This therefore implies that gas is being transported from outside corotation, a mechanism that can be explained in the framework of invariant manifolds, but not in the framework of resonant built rings. This constantly renewed supply of gas sustains star formation in the ring for extended periods of time even after the bar pushes all the gas to the centre and quenches star formation in the bar region. 

As the inner ring forms due to the bar, its star formation history could help reveal the formation time of the bar (similarly to how nuclear rings and discs can serve to age-date the bar -- see \citealt{Gadottietal2015,Gadottietal2019} and \citealt{BabaKawata2019}).
We see that, as expected, in the two cases with a prominent inner ring, the residual star formation in CR-bar is much higher than in those without an inner ring. Since the inner rings form after the bar, the star formation history of the ring could provide a method for determining the formation time of the bar. We explore this in the bottom panels of Figure \ref{fig:sfhring} where we show the SFR inside corotation minus that inside the bar (i.e. CR-bar) divided by the SFR in the bar. In these panels the formation of the bar is marked by the vertical dot-dashed line. In the galaxies without a ring this ratio is low, since almost all star formation which happens inside corotation takes place inside the bar region. However, when an inner ring forms this ratio increases above one and continues to rise, since while star formation quenches in the bar due to gas depletion, it is still ongoing in the ring (because as discussed above, manifolds can transport gas from outside corotation into the ring). While in the bottom panels of Figure \ref{fig:sfhring} the sudden increase in SFR does not mark the exact time of bar formation, it does provide a lower limit to the age of the bar.

\subsection{Effect of the bar on disc phase-space}
\label{sec:phasespace}

The second Gaia data release \citep{GaiaCollaboration2018} recently revealed a plethora of complex substructure in phase-space in the disc of the Milky Way, with one particularly prominent feature being the ridges observed in the space of tangential velocity, $V_{\phi}$ versus galactocentric radius $r$ \citep{Kawataetal2018,Antojaetal2018} and the correlation of these ridges with undulations in radial velocity $V_r$ \citep{Fragkoudietal2019}. 
In \cite{Fragkoudietal2019} we showed, using an N-body simulation of a Mikly Way-type galaxy, that these ridges and undulations can be the product of bar-induced resonances, and specifically that the OLR will create the largest ridge observed in this plane, with a prominent inwards and outwards moving $V_r$ component associated to it. This occurs due to the underlying resonant orbital structure at the OLR, where there are overlapping anti-aligned $x_1(1)$ and $x_1(2)$ orbits (see \citealt{Dehnen2000} and Figure \ref{fig:orbsAu18} for examples of these types of orbits in our fiducial model Au18).
Here we explore the effects of the bar on the disc kinematics in our sample of Auriga galaxies, specifically on the $V_{\phi}-r$ plane, as well as the relation between the kinematic signatures of the OLR and ages and chemical abundances.

In Figure \ref{fig:vphir_all}  we show logarithmic density plots of the $V_{\phi}-r$ plane in the discs of the five models explored in this study. To construct the plots we select all stars within 0.5\,kpc from the plane of the galaxy. We see that there are a number of ridges present in this plane in all cases, with one particularly prominent ridge present in most cases. The dashed line in each panel corresponds to the angular momentum of a circular orbit at the OLR radius. This line is associated to OLR resonant stars, as shown in Figure \ref{fig:resonancesAu18} where we carry out a spectral orbital analysis of stars in our fiducial model, Au18. We see that in all cases the OLR resonance is associated with the largest ridge in the $V_{\phi}-r$ plane, with the exception of Au13; in this case, the disc of the galaxy ends at around the OLR radius (this can be seen also by examining the top row of Figure \ref{fig:rgball}). We therefore see that the longest ridge in the $V_{\phi}-r$ plane being associated to the OLR is a generic feature apparent in a number of models, and can therefore be used as an independent method for estimating the location of the bar OLR, both for the Milky Way, as well as for external galaxies. 

In Figure \ref{fig:vphir_au18_agefe} we show the $V_{\phi}-r$ plane of Au18 with mass-weighted mean age (left), metallicity (middle) and $\alpha$-abundance (right) colour-coded (see Figure \ref{fig:vphiall} for all models). As can be seen in the Figure, all ridges, and especially the OLR ridge which is the most prominent, has on average younger stars associated to it. This could be due to the fact that colder populations are most affected by the resonances caused by the bar, or due to preferential ongoing star formation in these regions. Correspondingly, the ridge region also has on average higher metallicity [Fe/H] and lower mean alpha-abundances [$\alpha$/Fe]. Therefore, we find that the ridges in density are also apparent as ridges in age, metallicity and $\alpha$-abundance space. This has been shown to be the case also for the ridge structure of the Milky Way (see \citealt{Khannaetal2019} who combined the kinematic information on the ridges from Gaia DR2 with information on chemistry from the GALAH survey). There is therefore a plethora of information to be distilled by combining kinematics with chemistry in order to disentangle the origin of the different ridges seen in phase-space in the Milky Way, and we will explore this in more detail in upcoming work.

\section{Discussion}
\label{sec:discussion}

\subsection{The metal-poor vs metal-rich Milky Way bar/bulge}

In the Auriga simulations we find that bars and b/p's are predominantly metal-rich (see Figures \ref{fig:xyall_ages} and \ref{fig:xy_edgeon_feh_all}). This is in agreement with IFU spectroscopic studies of the inner regions of external galaxies, which find that a number of bars and b/p's are rather metal-rich, or as metal-rich as their surrounding disc (see e.g. \citealt{Pinnaetal2019,Gonzalezetal2017,Gadottietal2019,Neumannetal2020}). 
In this section we discuss these findings in the context of the inner few kpc of the Milky Way, which, as we will discuss below, appear to be rather metal-poor (including the bar/bulge region).

A number of surveys (e.g. GIBS, GES, APOGEE) which have explored the bulge of the Milky Way (i.e. inside $|l,b|<10\,\rm deg$) find that it has a significant metal-poor component at all latitudes, which leads to the Milky Way bulge being on average metal-poor, i.e. [Fe/H]<0 (e.g. see \citealt{Nessetal2013a,NessFreeman2016,Zoccalietal2017,RojasArriagadaetal2017,Fragkoudietal2018}). Furthermore, as discussed in previous sections, recently \citet{Bovyetal2019}, combining APOGEE DR16 with Gaia DR2 data with a machine-learning algorithm (see \citealt{Leungetal2019a,Leungetal2019b}) constructed `face-on' metallicity maps of the Milky Way. They find that the innermost regions of the Galaxy, i.e. the bar/bulge region, are metal-poor while the surrounding disc is more metal-rich. It is worth noting that the work of \citet{Bovyetal2019} is therefore consistent with the aforementioned surveys on the bulge of the Milky Way - i.e. they all find that the inner regions of the Milky Way are metal-poor (and see also the discussion in Section \ref{sec:effectring}). Therefore, it seems that the Milky Way might be somewhat an outlier with respect to other nearby barred galaxies, which tend to have metal-rich bar/bulge regions.

On the other hand, \citet{Weggetal2019} recently found that metal-rich stars in the bar region of the MW are on more elongated orbits, compared to metal-poor stars which are on rounder, more disc-like orbits. This behaviour was predicted in studies of discs with multiple stellar populations, in which `kinematic fractionation' occurs i.e. that metal-rich stars should be on more elongated orbits (e.g. \citealt{Debattistaetal2017,Fragkoudietal2017b}). This behaviour can also be explained with star formation occurring after the bar forms, which will preferentially place new stars on bar-like orbits. Based on this finding, and on a spatial separation of stars in the bar vs the inner disc, \citet{Weggetal2019} conclude that their findings are in tension with those of \citet{Bovyetal2019}.  

It is worth noting that kinematic `fractionation' in itself, is not in tension with a metal-poor bar-b/p region. As shown in \citet{Fragkoudietal2018}, in Section 5 (see specifically Figure 13), the overall metallicity of the inner bar/bulge region can be low if the metal-poor thick disc dominates at these radii (i.e. there will be a higher density of metal-poor stars in that region). This will occur if the thick disc is relatively massive, and has a shorter scale-length than the metal-rich thin disc, as is thought to be the case for the Milky Way (e.g. \citealt{Bensbyetal2011}). Since the model of \citet{Fragkoudietal2018} contains the effects of `kinematic fractionation', we see that this is not inconsistent with a metal-poor inner region inside the bar. Furthermore, as seen in Figure 13 of \citet{Fragkoudietal2018}, the end of the bar will be more metal-rich than the innermost region, a trend which is also found in \citet{Bovyetal2019} and consistent with the findings of \citet{Weggetal2019}.

Therefore, there seem to be two scenarios that can explain the apparently metal-poor bar/bulge of the Milky Way: the first one is that the Galaxy has a metal-poor inner region on average\footnote{Although see also \citealt{Schultheisetal2019} who find that the innermost degree of the MW – which contains the nuclear star cluster – is metal-rich.}, and therefore is perhaps an outlier compared to other local barred spiral galaxies (but see also \citealt{Zhuangetal2019}, who find that late-type spirals in the CALIFA survey can have positive metallicity gradients). In this case, the metal-rich inner disc of the galaxy can be explained via a thin+thick disc scenario with different scale-lengths \citep{Fragkoudietal2018}  in combination with a metal-rich inner ring formed due to the presence of the bar (as shown in Section \ref{sec:effectring}). The second scenario, is that the Milky Way has a bar/bulge region which is more metal-rich than the inner disc, but, due to some selection effects, a significant fraction of metal-rich stars are missing from the aforementioned surveys. In either case, we reiterate that having a metal-poor bar/bulge region is in fact consistent with having metal-rich stars on more elongated orbits compared to the metal-poor ones, since what sets the \emph{overall} metallicity of a region is the local density of metal-rich vs metal-poor stars, which is set by a combination of the mass and scale-lengths of these stellar populations.

\section{Conclusions}
\label{sec:summary}

In this paper, the first in a series exploring the properties of barred galaxies in the Auriga magneto-hydrodynamical cosmological zoom-in simulations, we focus on the Auriga galaxies which form prominent boxy/peanut (b/p) bulges by $z=0$. We explore their chemodynamical properties, comparing these to the properties of the Milky Way bar and bulge, thus allowing us to place constraints on the formation history of the Galaxy. We also examine the effects of bars on the inner and outer disc of their host galaxy, exploring how they redistribute stars and gas in inner rings and phase-space ridges. Our results are as follows:
\begin{itemize}
\item {\bf{\emph{Statistical properties:}}} We find that the Auriga suite of simulations reproduces well the fraction of barred galaxies as a function of redshift, as well as the properties of bars at $z=0$ as compared to observations (see also \citetalias{BlazquezCaleroetal2020}). The b/p's have a range of formation times, from 1 to 8\,Gyrs ago and can undergo multiple buckling phases, however their fraction at $z=0$ is lower than that found in observations (see Section \ref{sec:bpsample}).

\item {\textbf{\emph{Ages and abundances:}}} The face-on and edge-on distribution of ages and abundances are significantly affected by the bar and b/p, which redistribute stars according to the kinematic properties of the underlying stellar population (see e.g. \citealt{Fragkoudietal2017b,Athanassoulaetal2017} -- this process was dubbed `kinematic fractionation' in \citealt{Debattistaetal2017}.) This leads to age and abundance gradients along the bar minor axis, in which younger stars cluster at the ends and of the bar and along its major axis (see also \citealt{Neumannetal2020}). Also, the b/p's show an X-shaped age and abundance distribution, in which younger and more metal-rich stars trace the shape of the peanut. All the b/p's in our sample contain a significant fraction of stars younger than 5\,Gyrs, $\sim$30\% for our fiducial Milky Way model, Au18 (see Section \ref{sec:ageabund}).

\item {\textbf{\emph{Chemo-morphological properties:}}} Stellar populations in the bar and b/p show signs of `kinematic fractionation', i.e. younger and more metal-rich populations are trapped on more elongated bar-like orbits in the face-on projection, and have more prominent peanut shapes in their egde-on projection. This is a consequence of the amount of angular momentum lost by stellar populations with different kinematic properties, i.e. younger/colder stellar populations which are present in the disc before the bar forms lose more angular momentum compared to older/hotter populations. Populations born inside the bar region after bar formation do not have as much angular momentum to lose to begin with, because stars are born on elongated bar-like orbits (see Section \ref{sec:chemorph}).

\item {\textbf{\emph{Global kinematic properties:}}} When viewed edge-on the b/p's in our sample exhibit cylindrical rotation as well as X-shaped dispersion profiles (i.e. the peanut region has lower velocity dispersion). When viewed face-on, asymmetric b/p's (i.e. bars which are currently buckling) display a butterfly pattern in $V_z$, confirming the results of \citealt{Lokas2019}. This kinematic signature of buckling bars can help identify asymmetric b/p's which are viewed face-on (see Section \ref{sec:kin1} and Appendix \ref{sec:Appendixvels}).

\item {\textbf{\emph{Chemo-kinematics \& formation history:}}} We compare the chemo-kinematic properties of stellar populations in different metallicity bins in our b/p's to those of the Milky Way bulge (with the ARGOS survey -- \citealt{Nessetal2013b}). The haloes which best reproduce the kinematics of the Milky Way bulge, i.e. which have a rotating metal-poor component with a flat velocity dispersion profile, are Au17 \& Au18. These galaxies have the most quiescent merger histories of the entire Auriga sample of galaxies; their last major merger, i.e.  with stellar mass ratio $f_{\star}>0.25$, occurs $> 12\, \rm Gyrs$, and all subsequent mergers are $f_{\star}<0.05$ prograde or radial mergers. This suggests a stellar mass ratio of $\sim 1:20$ for the recently proposed Gaia Sausage/Enceladus merger (see e.g. \citealt{Haywoodetal2018,Helmietal2018,DiMatteoetal2018,Belokurovetal2019} and see Sections \ref{sec:kin2} and \ref{sec:SFH}).

\item{\textbf{\emph{Relation between morphology and kinematics:}}} The models which best reproduce the chemo-kinematics of the Milky Way bulge (Au17 \& Au18), i.e. which have fast rotating metal-poor components, also have metal-poor components with flattened density distributions in their edge-on projection, compatible with a thick disc distribution (see Sections \ref{sec:chemorph} and \ref{sec:kin2}).

\item {\textbf{\emph{Formation histories \& ex-situ fractions:}}} While the galaxies in our sample have diverse formation histories they all have low fractions of ex-situ material in their central regions (and see also \citetalias{Gargiuloetal2019}). The haloes with the most violent merger histories have larger ex-situ fractions (10\% for Au26) while those with the most quiescent merger histories (Au17 \& Au18) have ex-situ fractions of less than 1\%. Therefore, the two most MW-like b/p's, Au17 \& Au18 are essentially entirely made of in-situ stars. When considering only stellar populations with [Fe/H]<-1, the ex-situ fraction increases, but is still rather low for the two MW-like b/p's -- 13\% for Au17 and 37\% for Au18. The mean ex-situ fraction of stars for all Auriga galaxies with b/p's is 3\%, while for Auriga galaxies which do not form a b/p nor a bar it is 11\% and 18\% respectively - i.e. galaxies with b/p bulges tend to have lower ex-situ fractions overall, compared to galaxies without b/p's and without bars (see Section \ref{sec:SFH}). 

\item {\textbf{\emph{Inner rings:}}} Au18 and Au23 form an inner ring around the bar, which is star-forming and metal-rich. Such an inner ring, in combination with a thin+thick disc scenario, could explain the very metal-rich `inner disc' recently reported for the Milky Way \citep{Bovyetal2019}. Indeed the Milky Way is thought to harbour a gaseous inner ring, identified as the near and far 3\,kpc arms \citep{vanWoerdenetal1957,DameThaddeus2008}. The inner rings in our models show indications of being formed due to invariant manifolds, where gas is transported from outside to inside corotation. As inner rings form after bars, their star formation histories can help obtain a lower limit on the age of the bar (see Sections \ref{sec:bareffectdisc} and \ref{sec:discussion}).

\item {\textbf{\emph{Effect of the bar on phase-space in the disc:}}} In all the haloes in our sample the longest ridge in the $V_{\phi}-r$ plane is related to the bar OLR resonance (confirming the results of \citealt{Fragkoudietal2019}). This could provide an independent method for determining the bar pattern speed both in the Milky Way and in external galaxies. We also find that the ridges in this plane are associated with higher metallicities, lower alpha-abundances and younger ages, compared to the surrounding disc phase-space (see Sections \ref{sec:phasespace} and Appendix \ref{sec:appendixB}).

\end{itemize}

To summarise, we find that the models in our sample which best match the properties of the Milky Way bulge, have an in-situ origin (with <1\% of stars in the bulge formed ex-situ). Their metal-poor (-1<[Fe/H]<-0.5) populations rotate almost as fast as the more metal-rich populations ([Fe/H]>-0.5), and have have flattened morphologies, compatible with a thick disc. This is in agreement with recent chemodynamical studies carried out using tailored, isolated N-body simulations, in which the bulge of the Milky Way is composed of thin and thick disc populations (see e.g. \citealt{DiMatteo2016,Debattistaetal2017,Fragkoudietal2017b,Fragkoudietal2017c,Fragkoudietal2018}). 

Furthermore, contrasting the chemodynamical properties of the b/p's in our models with those of the Milky Way, allow us to place constraints on the merger history of the Galaxy, including on the recently proposed Gaia Sausage/Enceladus merger \citep{Belokurovetal2018,Haywoodetal2018b,Helmietal2018}.
One of our best-fitting models, Au18, experienced its last significant merger at $t_{\rm lookback}=9\,\rm Gyrs$, with the merging progenitor having a stellar mass ratio of 1:20 --  broadly in agreement with recent estimates of the Gaia Sausage/Enceladus merger (e.g. \citealt{Helmietal2018,DiMatteoetal2018,Belokurovetal2019}).
While our study does not involve an exploration of the full parameter space of merger times, mass ratios and orbital configurations, our results point to the Galaxy's largely quiescent merger history, where the last major merger ($f_{\star}\geq0.25$) took place at $z\geq3.5$, with only prograde or radial mergers with stellar mass ratio $f_{\star}\leq0.05$ occuring since $t_{\rm lookback}\sim12\,\rm Gyr$; more recent massive mergers would disturb the rotationally supported kinematics of the metal-poor populations in the bulge, thus not allowing to reproduce the Milky Way bulge's chemodynamical properties. 
We therefore see that with a diverse sample of Milky Way-type galaxies formed in the full cosmological context we can disentangle the effects of different formation mechanisms on the chemodynamical properties of bars and b/p bulges, shedding light on the formation history of the Galaxy and the origin of its bulge.


\section*{Acknowledgements}
FF thanks Wilma Trick, Paola Di Matteo, Dimitri Gadotti and Misha Haywood for comments on earlier versions of the manuscript which greatly improved its clarity, and for many interesting discussions. The authors thank the anonymous referee for a constructive report. The authors thank David Campbell and Adrian Jenkins for generating the initial conditions and selecting the sample of the Auriga galaxies, and Paola Di Matteo for the isolated N-body simulations used in Appendix B.
FM acknowledges support through the Program ``Rita Levi Montalcini'' of the Italian MIUR. 
I.G. acknowledges financial support from CONICYT Programa Astronom\'{i}a, Fondo ALMA-CONICYT 2017 31170048.
AM acknowledges support from CONICYT FONDECYT Regular grant 1181797.
FAG acknowledges financial support from CONICYT through the project FONDECYT Regular Nr. 1181264. FAG, AM and IG acknowledge funding from the Max Planck Society through a Partner Group grant.
This project was developed in part at the 2019 Santa Barbara Gaia Sprint, hosted by the Kavli Institute for Theoretical Physics at the University of California, Santa Barbara. This research was supported in part at KITP by the Heising-Simons Foundation and the National Science Foundation under Grant No. NSF PHY-1748958.
\bibliographystyle{mnras}
\bibliography{References}

\begin{appendix}
\section{Additional plots for all haloes}

\begin{figure*}
\centering
\includegraphics[width=0.8\textwidth]{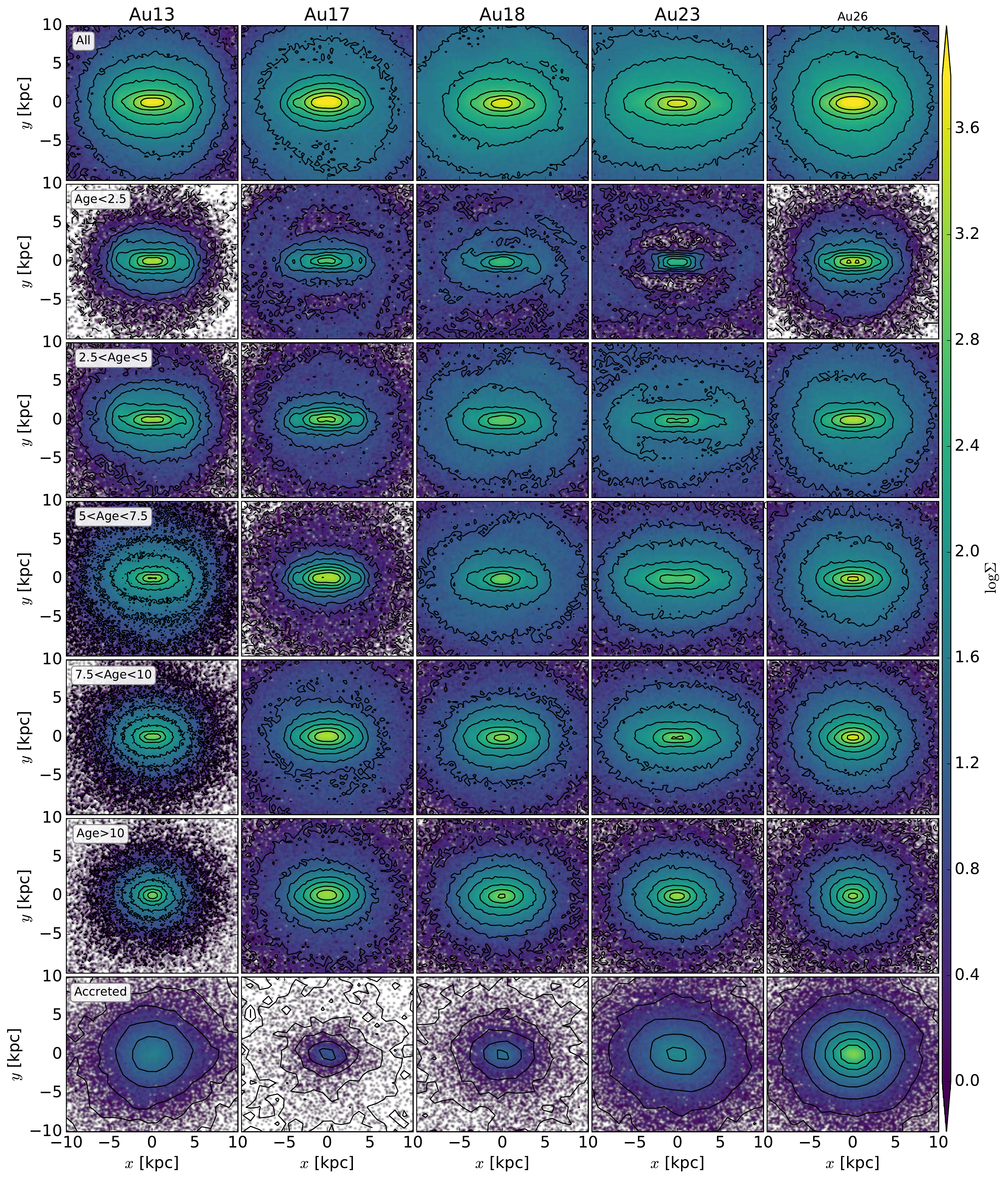}
\caption{\emph{Top row:} Face-on surface density for all stars within $|z|<3\,\rm kpc$ of the four haloes in our sample; the halo number is given at the top of each column. \emph{Subsequent rows}: Surface density of the stars in different age bins as denoted in the top left corner of the first column.} 
\label{fig:xy_faceon_ages_all}
\end{figure*}

\begin{figure*}
\centering
\includegraphics[width=0.8\textwidth]{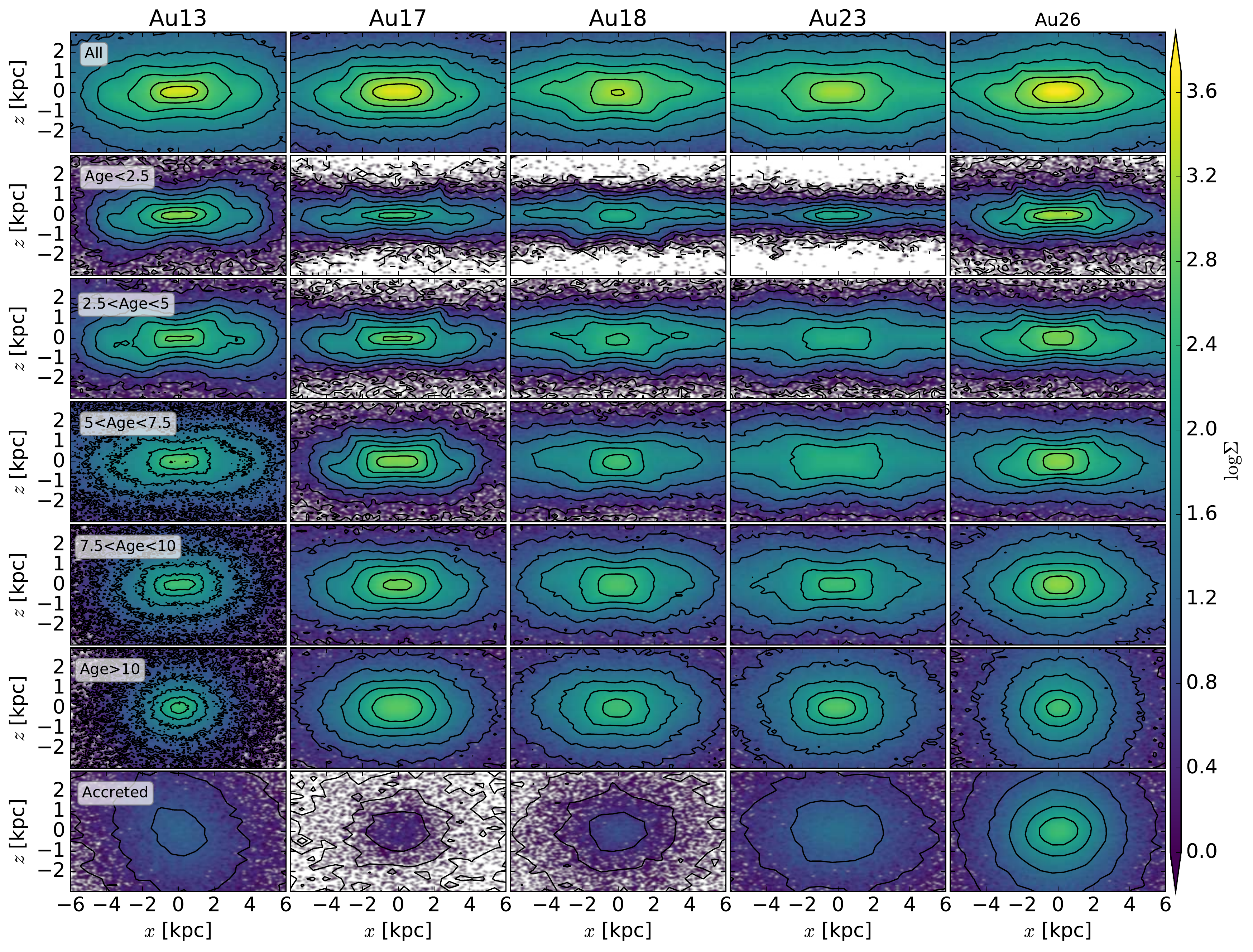}
\caption{Edge-on surface density projection for stars of different ages inside $r<4\,\rm kpc$ for the five haloes in this study. The ages are denoted in teh upper left corner of each panel. We see that the boxy/peanut morphology is more pronounced in the younger populations, while older populations appear rounder.} 
\label{fig:xy_edgeon_ages_all}
\end{figure*}

\begin{figure*}
\centering
\includegraphics[width=0.8\textwidth]{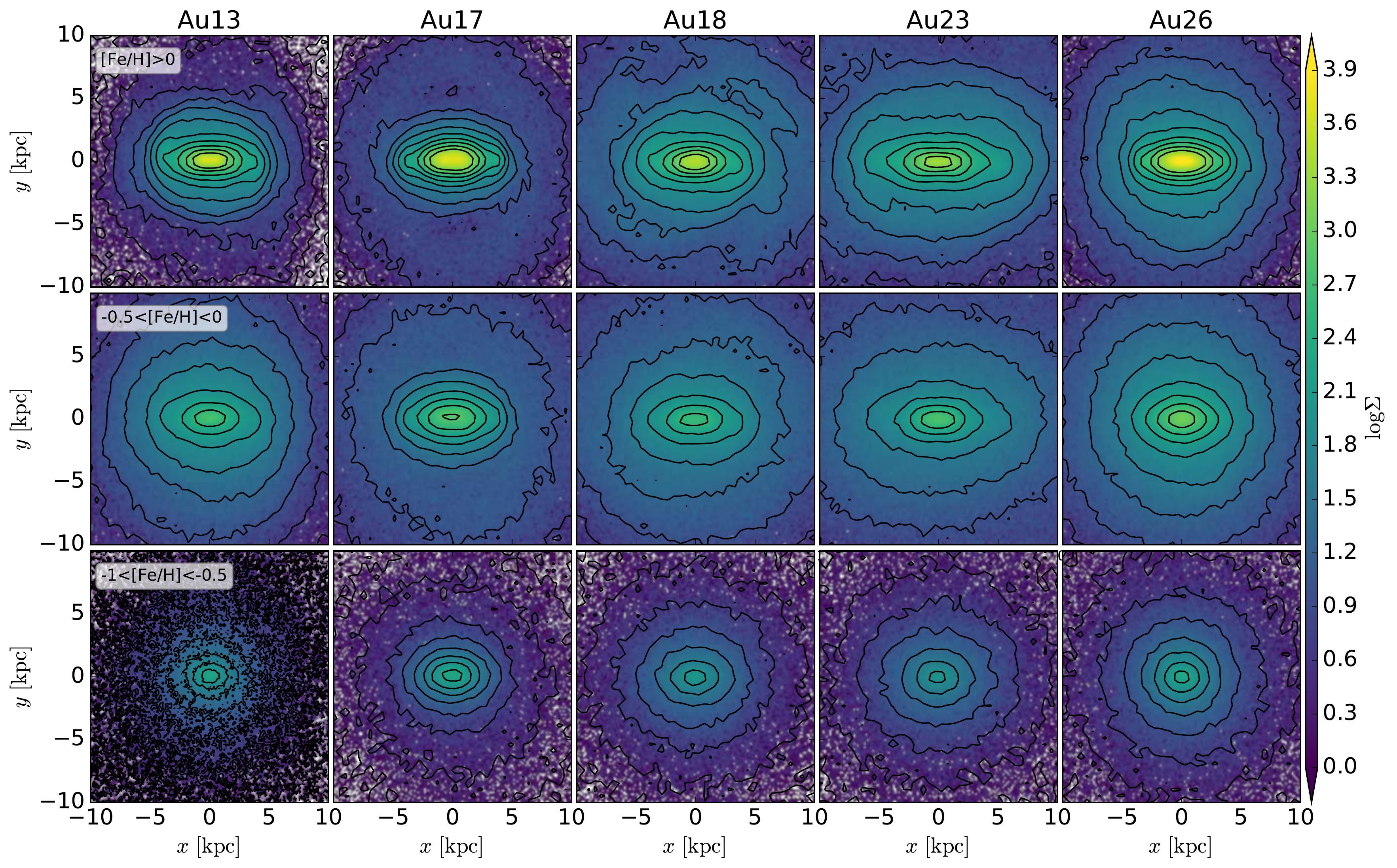}
\caption{Face-on surface density for three different metallicity bins, as denoted in the top left corner of the first column.} 
\label{fig:xy_faceon_feh_all}
\end{figure*}

Here we show additional plots of the chemo-morphological and chemo-kinematic properties for all haloes in our sample.
In Figures \ref{fig:xy_faceon_ages_all} -- \ref{fig:xy_faceon_feh_all} we show the face-on and edge-on surface density projections of the galaxies in our sample for different mono-age and mono-metallicity populations. In the top rows of Figures \ref{fig:xy_faceon_ages_all} and \ref{fig:xy_edgeon_ages_all} all stars are included, while in subsequent rows we show surface density projections for progressively older populations. The ages and metallicity of stars are marked in the top left corner of the first column. 

\begin{figure*}
\centering
\includegraphics[width=0.95\textwidth]{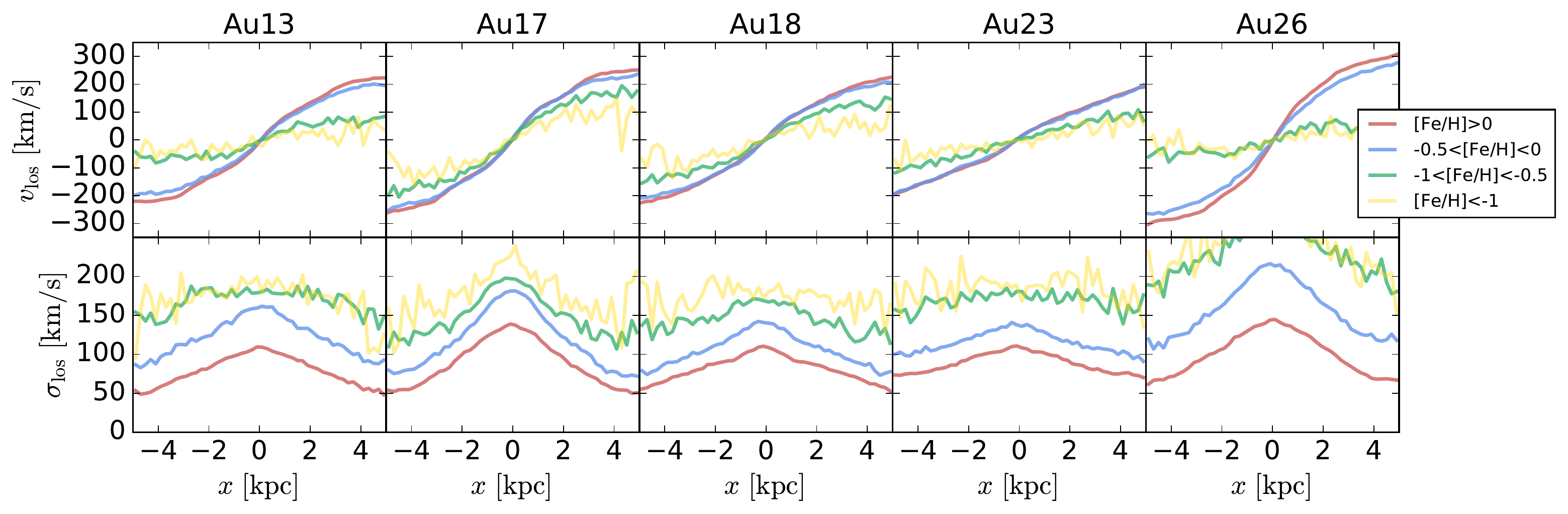}
\caption{Chemo-kinematic properties of stellar populations in the inner region of the five galaxies explored in this study: Line of sight velocities (top panels) and velocity dispersion (bottom panels) along the bar major axis for stars close to the plane ($|z|<0.5\,\rm kpc$). Stars are separated into four metallicity bins as denoted in the rightmost panel. We see that for haloes Au17 and Au18 the metal-poor component (-1 < [Fe/H] < -0.5) rotates at approximately the same velocity as the more metal-rich components, while for haloes Au13, Au23 and Au26 the metal-poor component has lower rotation. This is a consequence of the merger histories of these galaxies.} 
\label{fig:vlos_sigmalos_all_append}
\end{figure*}

In Figure \ref{fig:vlos_sigmalos_all_append} we show the line of sight velocity and velocity dispersion for the five galaxies in our sample as a function of the $x$-axis, where the bar is rotated to be along the $x$-axis. This is similar to Figure \ref{fig:morph_edgeon_abund}, but without normalising to the Milky Way mass and without transforming to Galactic coordinates.

\begin{figure*}
\centering
\includegraphics[width=0.8\textwidth]{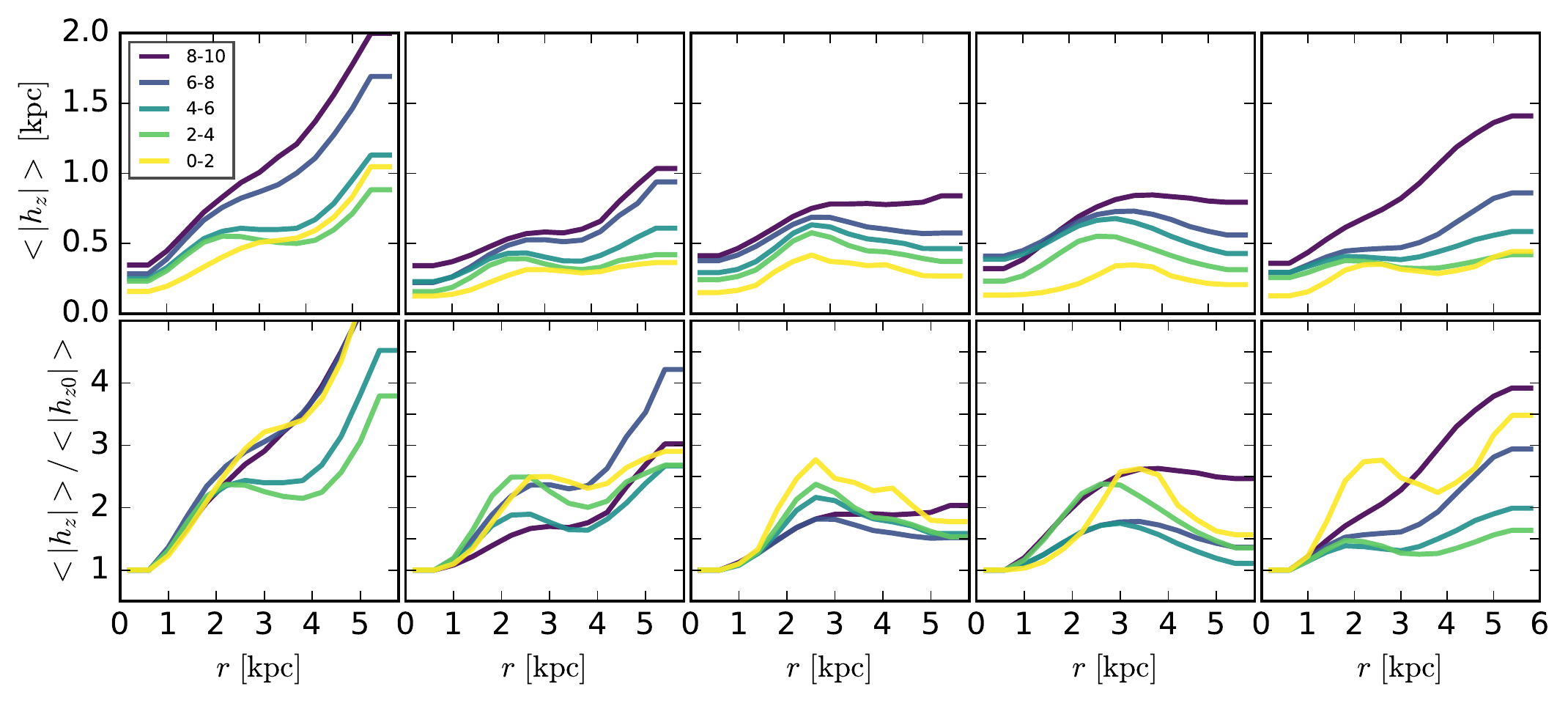}
\caption{\emph{Top:} Average height of mono-age populations (as indicated in the left panel) along the bar major axis. \emph{Bottom:} Average height of different mono-age populations along the bar major axis ($x$) normalised by the height in the centre for each mono-age population. We see that the younger populations exhibit a much more prominent peanut morphology and reach larger relative heights compared to hotter/thicker populations.} 
\label{fig:peanut_strength}
\end{figure*}

The stronger peanut shape of younger populations can be clearly seen in Figure \ref{fig:peanut_strength} where we show the average height $<|h_z|>$ -- normalised by $<|h_{z0}|>$, i.e. the scale-height in the centralmost region -- along the bar major axis for different mono-age populations (as indicated by the coloured lines in the top left corner of the first panel). We see that all populations flare towards the outer parts of the disc, and that in the region of the boxy/peanut bulge, the youngest populations tend to show the most prominent peanut shape, compared to the older populations (their relative increase in thickness at the location of the peanut is larger for the younger/thinner populations than for the older/thicker populations).

\begin{figure*}
\centering
\includegraphics[width=0.75\textwidth]{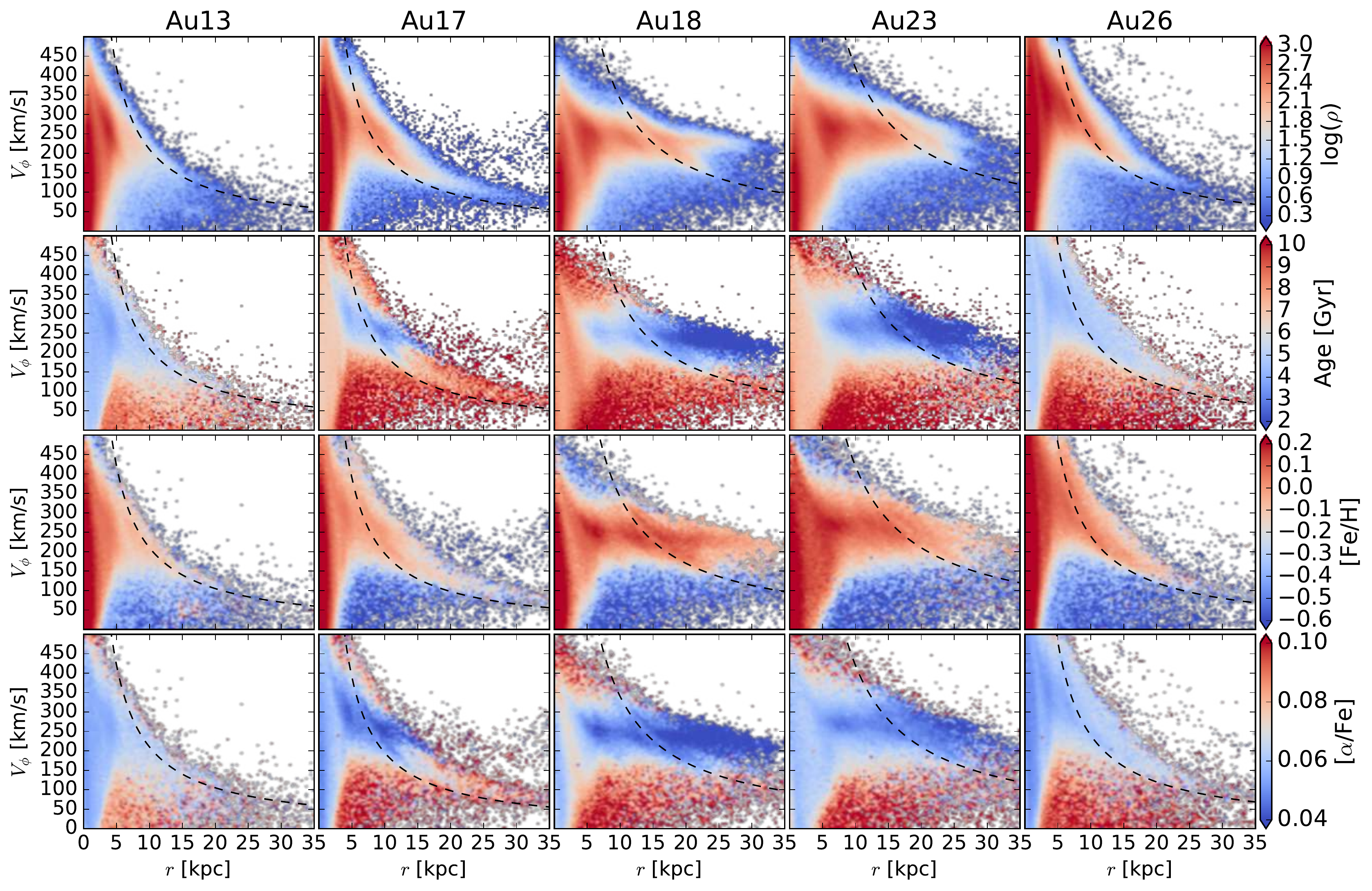}
\caption{Density, ages and abundances in the $V_{\phi}-r$ plane. \emph{Top row:} logarithmic density for each halo in the $V_{\phi}-r$ plane. \emph{Second row:} Average age in each pixel. \emph{Third row:} Average metallicity [Fe/H] in each pixel. \emph{Fourth row:} Average [$\alpha$/Fe] in each pixel. The dashed curve line indicates a line of constant angular momentum which passes through the OLR radius. We see that the largest ridge in the models corresponds to the ridge associated to the OLR.} 
\label{fig:vphiall}
\end{figure*}

In Figure \ref{fig:vphiall} we show the $V_{\phi}-r$ plane for all galaxies as well as the mean age (second row), [Fe/H] (third row) and $\alpha$-abundances (fourth row) in this plane colour coded.

\section{Kinematic signature of buckling bars}
\label{sec:Appendixvels}

\begin{figure*}
\centering
\includegraphics[width=0.75\textwidth]{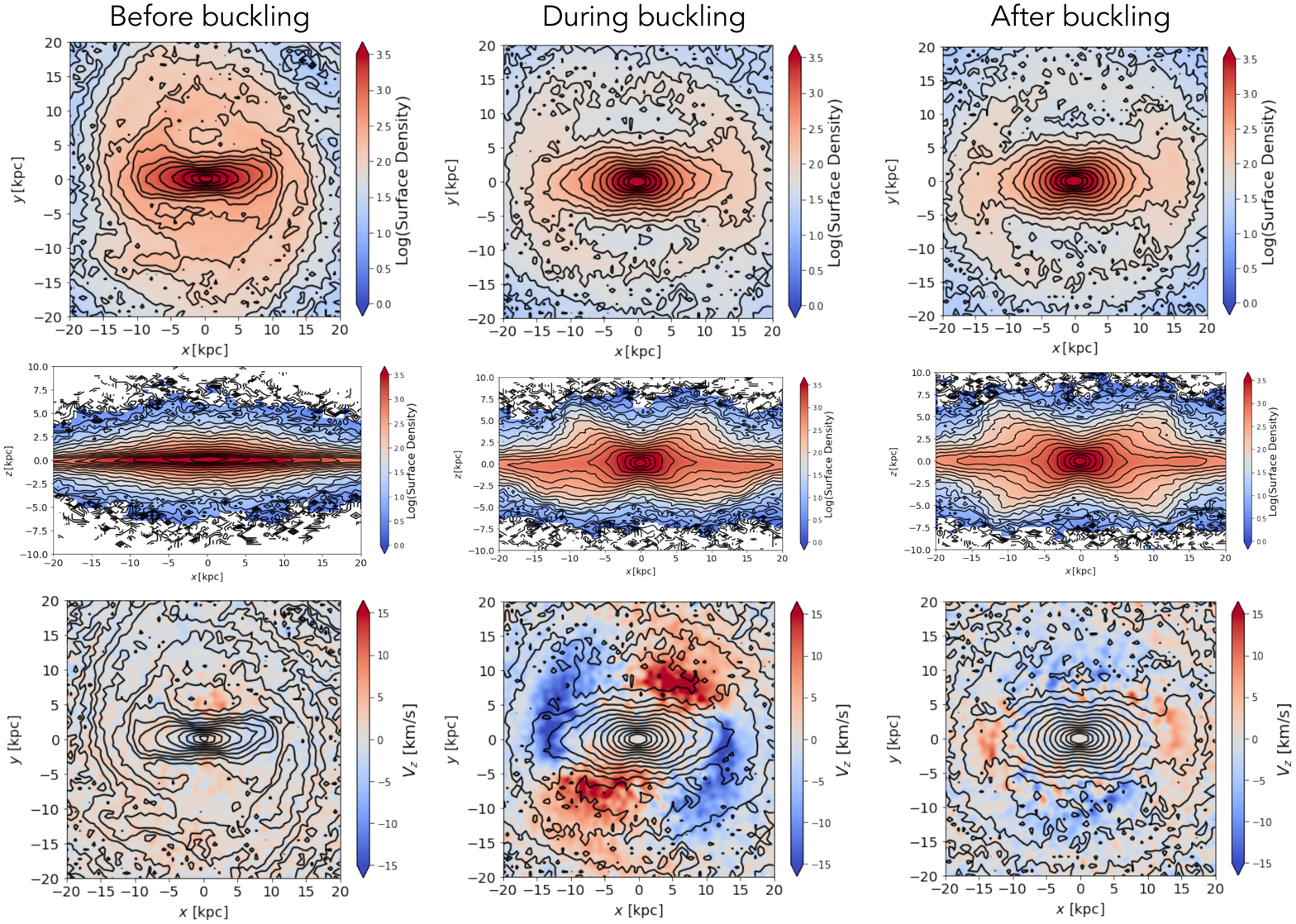}
\caption{Examining the kinematic signature of buckling bars: We use the isolated disc simulation presented in \citet{Fragkoudietal2017b} to explore the butterfly feature in $V_z$ and its relation to the asymmetric buckling phase of the bar. In the top rows we show the face-on surface density projection of the disc of the model, in the second row we show the edge-on surface density projection and in the bottom row we show projected $V_z$ when the disc is viewed face-on. We see that the butterfly feature in $V_z$ does indeed only appear during the buckling phase (middle columns), while before (left columns) and after (right columns) the buckling, there is no butterfly pattern in $V_z$.} 
\label{fig:isotestbuck}
\end{figure*}

Here we explore the kinematic signature of the asymmetric (or buckling) b/p's presented in Section \ref{sec:kin1}. To verify that this signature is due to the asymmetric b/p and not due to other external processes affecting discs in cosmological simulations (such as e.g. disc bending, interactions with satellites etc.) we use the isolated disc galaxy model presented in \citet{Fragkoudietal2017b}. The model is a purely collisionless simulation of an isolated Milky Way-type galaxy, with 1$\times 10^6$ particles in the disc and 5$\times 10^5$ particles in the dark matter halo (for details of the model we refer the reader to \citealt{Fragkoudietal2017b}). 

We show in Figure \ref{fig:isotestbuck} three snapshots from this model, one before the bar buckles (left), one during the buckling (middle) and one after the buckling (right panels). We see that the butterfly signature in $V_z$ is only present during the asymmetric buckling phase, and is present neither before the b/p formation, nor after the b/p buckles when it becomes symmetric. We therefore see that this signature is indeed inherent to an asymmetric b/p, due to the asymmetric orbital structure in the b/p.

\section{Bar-induced resonances \& orbits} 
\label{sec:appendixB}

\begin{figure*}
\centering
\includegraphics[width=0.8\textwidth]{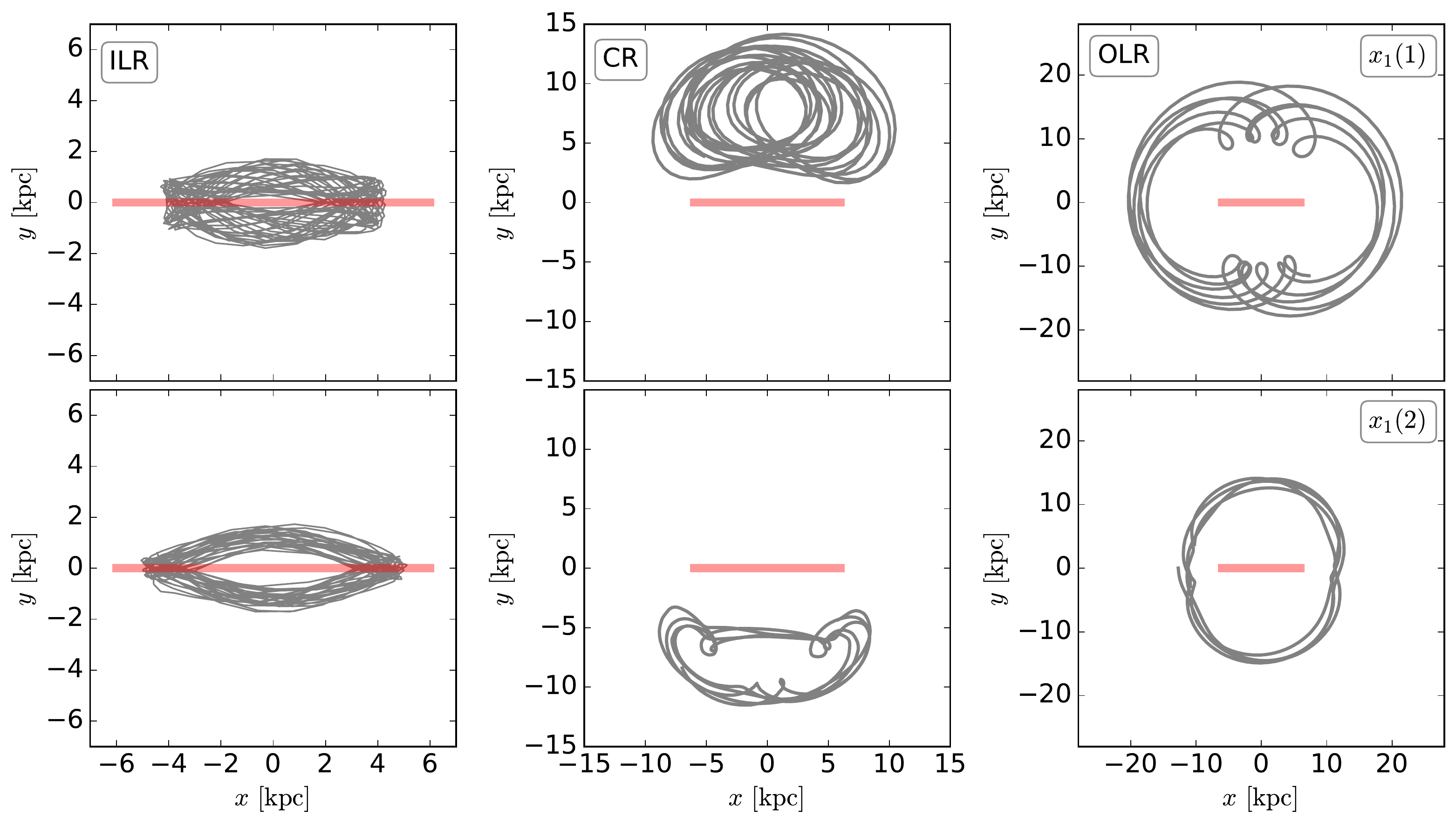}
\caption{Examples of orbits taken directly from the fiducial simulation Au18. In all panels the length and orientation of the bar is denoted with the red line. Note the different axis scales for each row. \emph{Left:} ILR resonant orbits. \emph{Middle:} CR resonant orbits. \emph{Right:} OLR resonant orbits with examples of the $x_1(1)$ family on the top (aligned with the bar) and the $x_1(2)$ family on the bottom (anti-aligned with the bar). These two families are responsible for the elongated ridge seen at the OLR radius.} 
\label{fig:orbsAu18}
\end{figure*}

\begin{figure*}
\centering
\includegraphics[width=0.8\textwidth]{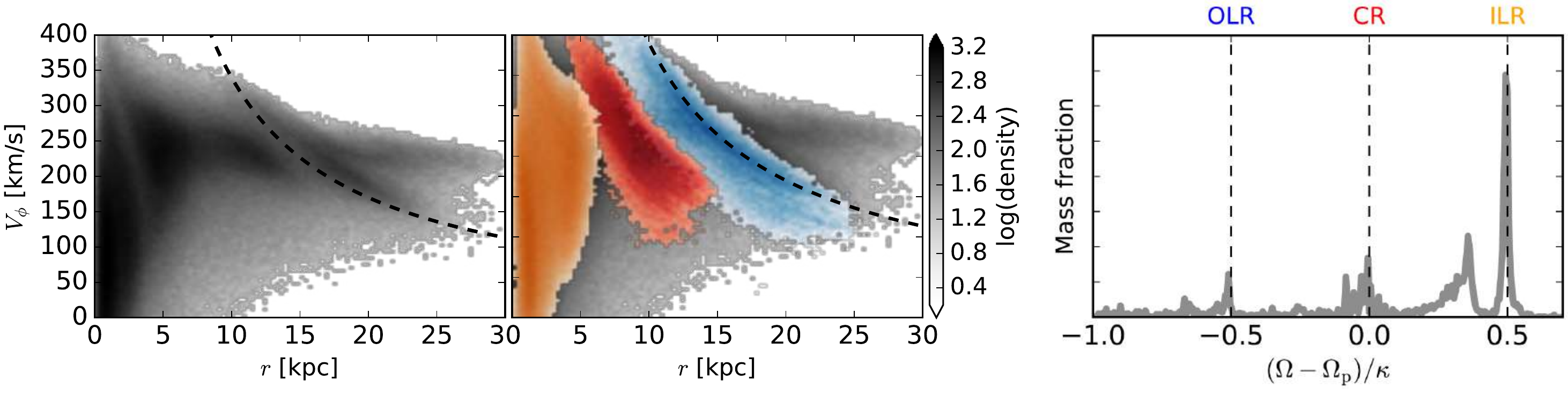}
\caption{Resonant structure of the $V_{\phi}-r$ plane. \emph{Left:} A logarithmic density plot of the $V_{\phi}-r$ plane of fiducial model Au18, with the OLR constant angular momentum line marked. \emph{Middle:} The same as in the left panel but with the ILR resonant (orange), CR resonant stars (red) and OLR stars (blue) marked. These stars are obtained by taking stars within an interval of $\pm 0.1$ of the resonant lines of the plot on the right. \emph{Right:} Frequency histogram of $10^4$ randomly selected stellar particles in the disc of the fiducial model Au18.} 
\label{fig:resonancesAu18} 
\end{figure*}

To explore the resonant orbital structure in our fiducial model we carried out a spectral analysis of $10^4$ stars in the disc, deriving their angular, radial and vertical frequencies ($\Omega$, $\kappa$ and $\nu$ respectively). We show the angular to radial frequency ratio of these stars as compared to the bar angular frequency ($\Omega_p$) in the rightmost panel of Figure \ref{fig:resonancesAu18}. The three main resonances due to the bar, the OLR, CR and ILR are marked with the vertical dashed lines. There are also additional peaks in the histogram which indicate higher order resonances, such as the broad peak around 0.36, which contains 1:3 orbits as well as higher order and irregular orbits. In Figure \ref{fig:orbsAu18} we show examples of these resonant orbits, where the bar length and orientation is marked with a red line (note the different scales of these figures). We see that in our model we can observe the typical $x_1$ bar-supporting orbits (top row), the horse-shoe like corotation orbits (second row) and the $x_1(1)$ and $x_1(2)$ type orbits found at the OLR (third row). To see where these star particles lie on the $V_{\phi}-r$ plane we select those which lie in the interval of $\pm0.1$ of the three main resonances (dashed lines in right panel of Figure \ref{fig:resonancesAu18}) and overplot them on the $V_{\phi}-r$ plane as shown in the middle panel of Figure \ref{fig:resonancesAu18}. The orange colours correspond to particles carrying out ILR orbits, the red to corotation particles and the blue to OLR particles. We see that indeed the blue OLR particles fall on the longest ridge. The OLR can also be marked in this plane with a line of constant angular momentum, which corresponds to the angular momentum a particle on a circular orbit at the OLR would have (i.e. OLR radius $\times$ circular velocity at OLR).

\end{appendix}
\bsp	
\label{lastpage}
\end{document}